\documentclass[twocolumn, trackchanges]{aastex6}
\shortauthors{Plotkin et al.}

\begin{document}

\title{The 2015 Decay of the Black Hole X-ray Binary V404 Cygni: Robust Disk-Jet Coupling and a Sharp Transition into Quiescence}
\shorttitle{V404 Cygni Entering Quiescence}

\author{
R.~M.~Plotkin\altaffilmark{1}, 
J.~C.~A.~Miller-Jones\altaffilmark{1}, 
E.~Gallo\altaffilmark{2}, 
P.~G.~Jonker\altaffilmark{3,4},  
J.~Homan\altaffilmark{5}, 
J.~A.~Tomsick\altaffilmark{6},
P.~Kaaret\altaffilmark{7},  
D.~M.~Russell\altaffilmark{8}, 
S.~Heinz\altaffilmark{9}, 
E.~J.~Hodges-Kluck\altaffilmark{2},
S.~Markoff\altaffilmark{10},
G.~R.~Sivakoff\altaffilmark{11}, 
D.~Altamirano\altaffilmark{12}, 
J.~Neilsen\altaffilmark{5,13}
}
\altaffiltext{1}{International Centre for Radio Astronomy Research - Curtin University, GPO Box U1987, Perth, WA 6845, Australia; richard.plotkin@curtin.edu.au}
\altaffiltext{2}{Department of Astronomy, University of Michigan, 1085 South University Avenue, Ann Arbor, MI 48109, USA}
\altaffiltext{3}{SRON, Netherlands Institute for Space Research, Sorbonnelaan 2, NL-3584 CA Utrecht, the Netherlands}
\altaffiltext{4}{Department of Astrophysics/IMAPP, Radboud University, P.O.\,Box 9010, 6500 GL Nijmegen, the Netherlands}
\altaffiltext{5}{MIT Kavli Institute for Astrophysics and Space Research, 77 Massachusetts Avenue 37-582D, Cambridge, MA 02139, USA}
\altaffiltext{6}{Space Sciences Laboratory, 7 Gauss Way, University of California, Berkeley, CA 94720-7450, USA}
\altaffiltext{7}{Department of Physics and Astronomy, University of Iowa, Iowa City, IA 52242, USA}
\altaffiltext{8}{New York University Abu Dhabi, PO Box 129188, Abu Dhabi, UAE}
\altaffiltext{9}{Department of Astronomy, University of Wisconsin-Madison, Madison, WI 53706}
\altaffiltext{10}{Anton Pannekoek Institute for Astronomy, University of Amsterdam, 1098 XH Amsterdam, The Netherlands}
\altaffiltext{11}{Department of Physics, University of Alberta, 4-181 CCIS, Edmonton, AB T6G 2E1, Canada}
\altaffiltext{12}{Department of Physics and Astronomy, University of Southampton, Highfield SO17 IBJ, UK}
\altaffiltext{13}{Hubble Postdoctoral Fellow}

\newcommand{\nh}{N_{\rm H}}    
\newcommand{\cmtwo}{{\rm cm}^{-2}}  
\newcommand{\lxledd}{L_X/L_{\rm Edd}}  
\newcommand{\lledd}{L_{\rm bol}/L_{\rm Edd}}   
\newcommand{\ledd}{L_{\rm Edd}}   
\newcommand{\ergs}{{\rm erg~s}^{-1}}
\newcommand{\flux}{{\rm erg~s^{-1}~cm^{-2}}}
\newcommand{\cps}{{\rm c~s}^{-1}}
\newcommand{\mdot}{\dot{m}}
\newcommand{\Mdot}{\dot{M}}
\newcommand{\msun}{M_{\odot}}
\newcommand{\mstar}{M_{\star}}
\newcommand{\mbh}{M_{\rm BH}}

\newcommand{\lr}{L_{\rm R}}   
\newcommand{\lx}{L_{\rm X}}     
\newcommand{\lkev}{l_{\rm 2keV}}    
\newcommand{\luv}{l_{\rm 2500}}     

\newcommand{\smbh}{SMBH}   
\newcommand{\imbh}{mBH}      
\newcommand{\imbhagn}{mBH AGN}   
\newcommand{\bh}{black hole}            
\newcommand{\Bh}{Black hole}            
\newcommand{\xrb}{BHXB}

\newcommand{\src}{V404 Cygni}
\newcommand{\fvar}{F_{\rm var}}

\newcommand{\tdelay}{15.4}   
\newcommand{\tdelayErr}{4.0}
\newcommand{\jetsize}{3.0}    
\newcommand{\jetsizeErr}{0.8}  

\newcommand{\dateone}{July 11}
\newcommand{\datetwo}{July 15}
\newcommand{\datethree}{July 20}
\newcommand{\datefour}{July 23}
\newcommand{\datefive}{July 28}
\newcommand{\datesix}{July 30}
\newcommand{\dateseven}{August 1}
\newcommand{\dateeight}{August 5}

\newcommand{\note}[1]{\authorcomment1{#1}}

\begin{abstract}
We present simultaneous X-ray and radio observations of the black hole X-ray binary \src\ at the end of its 2015 outburst.  From 2015 \dateone--\dateeight\ we monitored \src\  with \textit{Chandra}, \textit{Swift}, and \textit{NuSTAR} in the X-ray, and with the Karl G. Jansky Very Large Array  and the Very Long Baseline Array in the radio, spanning a range of luminosities that were poorly covered during its previous outburst in 1989 (our 2015 campaign covers $2\times10^{33} \lesssim L_{\rm X} \lesssim 10^{34}~\ergs$).  During our 2015 campaign, the X-ray spectrum evolved rapidly from a hard photon index of $\Gamma \approx 1.6$ (at $\lx \approx 10^{34}~\ergs$) to a softer $\Gamma \approx 2$ (at $\lx \approx 3 \times 10^{33}~\ergs$).  We argue that  \src\  reaching $\Gamma \approx 2$ marks the beginning of the quiescent spectral state, which occurs at a factor of $\approx$3-4 higher X-ray luminosity than the average pre-outburst luminosity of $\approx 8\times10^{32}~\ergs$.    \src\ falls along the same radio/X-ray luminosity correlation that it followed during its previous outburst in 1989, implying a robust disk-jet coupling.  We exclude the possibility that a synchrotron cooled jet dominates the X-ray emission in quiescence, leaving synchrotron self-Compton from either a hot accretion flow or from  a radiatively cooled jet  as the most likely sources of X-ray radiation, and/or particle acceleration along the jet  becoming less efficient in quiescence.   Finally, we present the first indications of  correlated radio and X-ray variability on minute timescales in quiescence, tentatively measuring the radio emission to lag the X-ray by $15\pm 4$ min, suggestive of X-ray variations propagating down a  jet of length $<$3.0 AU.
 
\end{abstract}


\section{Introduction} 
\label{sec:intro}

After spending 26 years in quiescence, the low-mass black hole X-ray binary (\xrb ) \src\ was detected in outburst on 15 June 2015 \citep{barthelmy15, kuulkers15, negoro15, younes15}, prompting an array of multiwavelength observations.   For about two weeks \src\ displayed spectacular variability from the radio through the gamma-ray \citep[e.g.,][]{mooley15, motta15, rodriguez15, roques15, tetarenko15, trushkin15, gandhi16, jenke16, kimura16, marti16, walton16}, while also launching powerful outflows in the form of ballistic synchrotron jets \citep{tetarenko15a} and fast disk winds \citep{king15, munoz-darias16}.   During these two weeks, \src\ also contained a high column of absorbing material ($\approx10^{24}~\cmtwo$), likely caused by material expelled from the accretion disk obscuring the central engine \citep{motta16}.    On 2015 June 26, \src\ exhibited  the brightest X-ray flare of the outburst, which was followed by a sudden decrease in flux \citep{sanchez-fernandez16}, and the  source began a gradual decay back toward quiescence  at all wavebands \citep[e.g.,][]{ferrigno15, martin-carrillo15, tetarenko15b, walton15}.  The high column density during the initial two weeks dissipated and approached the pre-outburst value of $\approx$$10^{22}~\cmtwo$ by 2015 July 2-5 \citep{sivakoff15}.    

   This paper focuses on simultaneous X-ray and radio monitoring   of the tail end of the decay, from 2015 July 11 - August 5.   \src\ is one of our prime  laboratories for testing models of accretion physics against observations, largely because it is a nearby \xrb\ with an accurate distance of $2.39 \pm 0.14$ kpc measured from radio parallax \citep{miller-jones09}, allowing precise calculations of its energetics.  Furthermore, its orbital parameters are well-determined, with the system  comprised of  a $9.0^{+0.2}_{-0.6}~\msun$ black hole and a K3 III companion star  \citep{khargharia10} in a $6.473 \pm 0.001$ d orbit \citep{casares92}, with an inclination of $67^{+3}_{-1}$ deg \citep{khargharia10}.  

 During a typical outburst, \xrb s begin their final descent while in the ``low-hard state'' ($\lx \lesssim 10^{-2}~\ledd$; for \src, the Eddington luminosity $\ledd = 1.13 \times 10^{39}\ \ergs$), where their X-ray spectra can be described by a power-law with photon index\footnote{The photon index $\Gamma$ is defined such that the photon number density $N_E$ (per unit photon energy $E$) follows $N_E \propto E^{-\Gamma}$.}  
$\Gamma \approx 1.6-1.7$ \citep[see, e.g.,][for reviews on \xrb\ spectral states]{remillard06, belloni10}.   In the low-hard state, the inner regions of the disk are under-luminous because material is unable to efficiently cool via radiative losses \citep[e.g.,][]{ichimaru77, narayan94, abramowicz95, yuan14}, resulting in a hot,  geometrically thick, radiatively inefficient accretion flow (RIAF), which is likely to develop outflows \citep[e.g.,][]{narayan95, blandford99}.  Compact radio emission is nearly always associated with low-hard state \xrb s, which is usually interpreted as optically thick synchrotron radiation from partially self-absorbed relativistic jets \citep[e.g.,][]{blandford79, hjellming88, fender01}, and these jets may carry away a substantial fraction of the accretion power \citep[e.g.,][]{fender03}.    Finally,  there is evidence that a cool, thin disk \citep[e.g.,][]{shakura73} can persist close to the innermost stable circular orbit toward the bright end of the low-hard state \citep{miller06, reis10, uttley11}, but the disk is observed to  recede at lower luminosities \citep[see, e.g.,][]{tomsick09}.

As \xrb s  fade  from the low-hard state toward quiescence, their X-ray spectra become softer \citep[e.g.,][]{tomsick01, tomsick04, kalemci05, wu08, sobolewska11, armas-padilla13}   until they display $\Gamma \approx 2.1$ \citep[e.g.,][]{kong02, corbel06, plotkin13, armas-padilla14, reynolds14, yang15}.  From an ensemble of quiescent \xrb s, \citet{plotkin13} argue that the X-ray spectral softening completes by 10$^{-5}~\ledd$, at which point the X-ray spectral shape remains constant as \xrb s continue to fade \citep[also see][]{sobolewska11}.  However, the cause of the softening remains unclear, largely because observations with sufficient sensitivity and cadence to track the photon index as it approaches, reaches, and eventually saturates to $\Gamma \approx 2.1$ are  scarce (see, e.g.,  \citealt{kalemci05, homan13} for examples of the best covered decays so far).     

Combining X-ray spectral information with radio observations can yield insight into the cause of the X-ray spectral softening, and how \xrb s may differ between quiescence and the low-hard state.  During the decay toward quiescence, hard state \xrb s display correlated X-ray and radio variations on day to week timescales, such that individual systems travel along non-arbitrary paths through the radio/X-ray luminosity plane \citep{corbel13, gallo14}.  These correlations are  taken as evidence for couplings between the inner accretion flow (probed by X-rays) and the compact jet (probed by the radio), and the slope of radio/X-ray luminosity correlations places constraints on the physical mechanisms responsible for the observed radiation \citep[e.g.,][]{heinz03, markoff03}.  Yet, even among the systems with the best multiwavelength coverage of the decay toward quiescence \citep{jonker10, jonker12, ratti12}, none also contains high quality spectral information around the key parameter space of $\lx \approx 10^{-5} \ledd$, and it is observationally unclear how any motion through the radio/X-ray plane connects to changes in the X-ray spectrum.

 The 2015 outburst of \src\ offered a unique opportunity to obtain sensitive X-ray and radio spectral observations of a \xrb\ around $\lx \approx 10^{-5} \ledd$ as it transitions into quiescence.  Using primarily \textit{Chandra}, \textit{Swift}, and the Karl G.\ Jansky Very Large Array (VLA), we obtained  simultaneous X-ray and radio monitoring observations over three weeks.   Of key importance is the relatively small distance to  \src, which allows high signal-to-noise ($S/N$) observations at the desired luminosities.   Also, the long orbital period of \src\ implies a large accretion disk and high mass transfer rate from the companion star \citep[][and references therein]{menou99}, such that, when not in outburst, \src\ has the highest luminosity of any  known \xrb\ with a well-determined distance ($\lx \approx 10^{33}~\ergs$; \citealt{bernardini14}; an exception is the anomalously luminous, but distant,  \xrb\ GS $1354-64$; \citealt{reynolds11}).  Thus, \src\ is arguably the best-studied quiescent \xrb\  \citep[e.g.,][]{casares92, shahbaz94, narayan97, hynes04, hynes09,  miller-jones08, hynes09, bernardini14, bernardini16a, xie14, markoff15, rana16}, thereby allowing rich comparisons between our observations and its pre-outburst  properties.

In Section~\ref{sec:obs} we describe our observations and data reduction.  We present results in Section~\ref{sec:res}, which are discussed in Section~\ref{sec:disc}.   During its 26 years of quiescence between outbursts, \src\ displayed strong X-ray variability by up to a factor of 5-8 over its average X-ray flux (0.5-10 keV) of  $1.2 \pm 0.3 \times 10^{12}~\flux$\ (corresponding to $\lx = 8 \times 10^{32}~\ergs = 10^{-6.2}~\ledd$; \citealt[][note, we convert their reported 0.3-10 keV values to 0.5-10 keV here]{bernardini14}).  Throughout the paper, we refer to the above as the ``pre-outburst'' flux of \src.  At the beginning of Section~\ref{sec:disc}, we  argue  that \src\ enters the quiescent spectral state at 3-4 times higher luminosity.  Therefore,   throughout the paper we define quiescence for \src\ to correspond to $F_{\rm X} \lesssim 5\times10^{12}~\flux$ ($\lx \lesssim 3 \times 10^{33}~\ergs \approx 10^{-5.6}~\ledd$).   All uncertainties are quoted at the 68\% (1$\sigma$) level, unless stated otherwise.  We generally report X-ray fluxes and luminosities from 0.5-10 keV, except for  when we discuss radio/X-ray correlations when we adopt  1-10 keV luminosities to ease comparisons with the literature \citep[e.g.,][]{corbel13, gallo14}.

\begin{table}[htbp]
\begin{deluxetable*}{l c c c c c c c c c}
\tablecaption{Summary of \textit{Chandra}, \textit{Swift}, and VLA  Observations \label{tab:obslog}}
\decimals
\tablecolumns{10}
\renewcommand\arraystretch{1.2}
\tablehead{
		\multicolumn{2}{c}{}  &
		\multicolumn{5}{c}{\textit{Chandra}}  &
		\colhead{\textit{Swift}} & 
		\multicolumn{2}{c}{VLA} \\ 
                \colhead{MJD}      				&  
		\colhead{Date}  				&  
		\colhead{Start Time}  			& 
		\colhead{$t_{\rm exp}$}   			&  
		\colhead{ObsID}                		& 
		\colhead{Gratings}				& 
		\colhead{Subarray}       			& 
		\colhead{$t_{\rm exp}$}                      & 
		\colhead{Start Time}       			&   
		\colhead{$t_{\rm exp}$}  \\   		
		\colhead{}                            		& 
		\colhead{(2015)}         					& 
		\colhead{(UT HH:MM:SS)}		& 
		\colhead{(ks)}					& 
		\colhead{}       					& 
		\colhead{}						& 
		\colhead{}				 	&	
		\colhead{(ks)}		&  
		\colhead{(UT HH:MM:SS)}			& 
		\colhead{(ks)}                      
}
\colnumbers
\startdata
57214 &  \dateone\        & 13:02:49 &     39.3           &     17701         &  HETG  & None & 1.2 &  \nodata  &  \nodata   \\
57218  &  \datetwo\        & 03:22:13   &     3.7          &    16702         &  HETG  &  1/2 &    7.3  & 04:12:40   & 0.5 \\ 
57223 &  \datethree\        & 04:01:48  &    8.8          &     16703         &  HETG  &  None  & 10.9  & \nodata    &  \nodata   \\
57226\tablenotemark{a} &  \datefour\        &  08:36:58 &     19.3          &     16704         &  HETG  & None    &  8.2  &  11:29:56  & 0.5 \\ 
57231 &  \datefive\        &  09:42:05 &     17.9           &     16705         &  None    &   1/8 & 12.0  &  09:55:28   &  2.2 \\ 
57233 &  \datesix\        &  \nodata &     \nodata           &    \nodata         &  \nodata  & \nodata  &  \nodata &  09:47:36 & 2.2 \\ 
57235 &  \dateseven\        & 12:25:25  &     26.2           &     16706        &  None  & 1/8 &  11.7  & 11:24:28    & 5.3 \\ 
57239 &  \dateeight\        &  03:33:27 &     42.7           &     16707         &  None  & 1/8  &  9.7  &  03:33:00  & 5.6 
   \enddata
\vspace{0.3cm}
 \tablenotetext{a}{We also obtained a 40 ks \textit{NuSTAR} observation, taken between 2015 July 23 UT 08:21 - July 24 UT 11:01.}
\tablecomments{Column (1) modified Julian date of each observing epoch.
Column (2) UT date of each epoch.
Columns (3)--(7) pertain to the \textit{Chandra} observations, including the UT start times (column 3), the exposure time (column 4), the observation identification number (column 5), whether or not the HETG was in place to act as a pileup mitigation filter (column 6), and the subarray read from the  ACIS chip (column 7).  The \textit{Chandra} observation from \dateone\ was from a separate DDT (see Section \ref{sec:obs:epochone}).
Column (8) lists the total exposure times of the \textit{Swift} observations, which were scheduled as series of 1--2 ks snapshots over 8-19 hours on each date, with at least one snapshot overlapping with the \textit{Chandra} observations, except for on \dateone.  The \textit{Swift} observation on \dateone\ was from a single snapshot that started at UT 07:03  (see Section~\ref{sec:obs:epochone}).
 Columns (9)-(10) pertain  to the radio observations (VLA project  code SG0196), including the UT start times (column 9) and the total observing times on source (column 10).  A radio observation from \dateone\ was obtained from the VLBA over UT 07:03--07:28, simultaneous with \textit{Swift} on that date (but not with \textit{Chandra}; see Section~\ref{sec:obs:epochone}).  
 }
\end{deluxetable*}
\end{table}
\renewcommand\arraystretch{1}

 \section{Observations}
 \label{sec:obs}
 
We triggered joint \textit{Chandra}/VLA observations through a cycle 16 \textit{Chandra}  target of opportunity (ToO) program (proposal ID 16400196; PI Plotkin).  This program included six \textit{Chandra} observations  taken  with an approximately four day cadence over a three week period between 2015 July 15 -- 2015 August 5.    Exposure times generally increased over time, ranging from 4 -- 43 ks.     We also arranged for ToO observations with the X-ray Telescope \citep[XRT;][]{burrows05} onboard the \textit{Swift} satellite \citep{gehrels04} to coincide with each \textit{Chandra} epoch (PI Plotkin).   The XRT exposures ranged from 7 -- 12 ks, composed of individual 1--2 ks snapshots spread over 8--19 hours on each date, with at least one snapshot on each date   simultaneous with a portion of each \textit{Chandra} observation.   On \datefour, we also arranged for a simultaneous 40 ks observation with the Nuclear Spectroscopic Telescope Array (\textit{NuSTAR}; \citealt{harrison13}). 

Simultaneous VLA radio observations were scheduled for the beginning of each \textit{Chandra} observation (VLA exposures also generally increased over time,  but  with each VLA observation  being shorter than the corresponding \textit{Chandra} one).  The VLA did not obtain usable data on our second epoch (20 July).  To make up for that  epoch, the VLA scheduled an extra observation on 30 July, for which there was not any corresponding X-ray data.  Thus, we obtained  observations on six dates in each waveband, and a total of five epochs included periods of strictly simultaneous overlap.   We also considered a set of simultaneous X-ray and radio observations taken on \dateone, obtained through separate  programs (see Section~\ref{sec:obs:epochone}).  A summary of our observations appears in Table~\ref{tab:obslog}.

\begin{table*}[htbp]
\begin{deluxetable*}{l l l L L L L L L L L L }
\tablecaption{X-ray Properties \label{tab:xray}}
\decimals
\tabletypesize{\footnotesize}
\tablecolumns{12}
\tablewidth{0.9\textwidth}
\renewcommand\arraystretch{1.2}
\tablehead{
		\colhead{}        & 
		\multicolumn{5}{c}{\textit{Chandra}}    & 
		\multicolumn{5}{c}{\textit{Swift}}          & 
		\colhead{}  \\   
                \colhead{Date}      				&  
		\colhead{N$_{\rm tot}$}  			&  
		\colhead{N$_{\rm bkg}$}  			& 
		\colhead{Net  Count Rate}   		&  
		\colhead{$f_{\rm 0.5-10 keV}$}          & 
		\colhead{$L_{\rm 0.5-10 keV}$}		& 
		\colhead{N$_{\rm tot}$}  			&  
		\colhead{N$_{\rm bkg}$}  			& 
		\colhead{Net  Count Rate}   		&  
		\colhead{$f_{\rm 0.5-10 keV}$}            & 
		\colhead{$L_{\rm 0.5-10 keV}$}		 & 
		\colhead{$\sigma_{\rm sys}$}       \\        
		\colhead{(2015)}                            		& 
		\colhead{(cts)}         					& 
		\colhead{(cts)}						& 
		\colhead{(cts s$^{-1}$)}				& 
		\colhead{(10$^{-12}$ cgs)}       			& 
		\colhead{(10$^{33}$ cgs)}       			& 
		\colhead{(cts)}         					& 
		\colhead{(cts)}						& 
		\colhead{(cts s$^{-1}$)}				& 
		\colhead{(10$^{-12}$ cgs)}       			& 
		\colhead{(10$^{33}$ cgs)}       			& 
		\colhead{(10$^{-12}$ cgs)}       			 
}
\colnumbers
\startdata
July 11\tablenotemark{a}                  &                     3608 &                      2.2 & 0.092 \pm 0.002          & 14.0 \pm 0.2           & 9.6 \pm 1.1     &  314 &                      5.5 & 0.257 \pm 0.015          & 19.8 \pm 1.2           & 13.5 \pm 1.8                               & \pm 5.0    \\
July 15                  &                      241 &                      0.2 & 0.065 \pm 0.004          & 8.3 \pm 0.6     & 5.7 \pm 0.8    &  524 &                     14.0 & 0.070 \pm 0.003          & 7.6 \pm 0.3        & 5.2 \pm 0.7           & \pm2.6            \\
July 20                  &                      600 &                      0.2 & 0.068 \pm 0.003          & 9.6 \pm 0.4      & 6.6 \pm 0.8     & 1020 &                     20.5 & 0.092 \pm 0.003          & 11.0 \pm 0.4     & 7.5 \pm 0.9            & \pm 4.8           \\
July 23\tablenotemark{b}                 &                      673 &                      0.6 & 0.035 \pm 0.001          & 4.6 \pm 0.2      & 3.1 \pm 0.4      & 383 &                     15.5 & 0.045 \pm 0.002          & 4.5 \pm 0.2       & 3.1 \pm 0.4         & \pm 2.2            \\
July 28                  &                     3333 &                      1.4 & 0.186 \pm 0.003          & 4.7 \pm 0.1     & 3.2 \pm 0.4      & 531 &                     15.5 & 0.043 \pm 0.002          & 4.8 \pm 0.2       & 3.3 \pm 0.4          & \pm 2.7            \\
August 1                 &                     3802 &                      1.5 & 0.145 \pm 0.002          & 3.6 \pm 0.1    & 2.5 \pm 0.3      &  403 &                     19.0 & 0.033 \pm 0.002          & 3.4 \pm 0.2       & 2.3 \pm 0.3         & \pm 1.5            \\
August 5                 &                    13651 &                      2.4 & 0.320 \pm 0.003          & 8.4 \pm 0.1   & 5.8 \pm 0.7      & 801 &                     14.5 & 0.081 \pm 0.003          & 8.5 \pm 0.3      & 5.8 \pm 0.7          & \pm 4.8             \\
\enddata
\tablenotetext{a}{The \textit{Chandra} observation is from a DDT program.  The \textit{Swift} observation is not simultaneous with \textit{Chandra} (see Table~\ref{tab:obslog} and Section~\ref{sec:obs:epochone}). }
\tablenotetext{b}{From \textit{NuSTAR}, we obtained average count rates from 3-79 keV  of $0.049\pm 0.001$ and $0.045\pm 0.001$\,s$^{-1}$ for FPMA and FPMB, respectively.}
\end{deluxetable*}
\vspace{-1.2cm}
\tablecomments{  
Column (1) observation date. %
Columns (2)--(6) present information from the \textit{Chandra} observations.  Column (2) total number of counts in source aperture. %
Column (3) number of estimated background counts in source aperture from 0.5-10 keV. %
Column (4) net count rate. %
Column (5) model unabsorbed flux from 0.5-10 keV, in units of 10$^{-12}~\flux$. Errors represent statistical uncertainties. %
Column (6) model luminosity from 0.5-10 keV, in units of 10$^{33}~\ergs$.  Errors include the uncertainty on the distance to the source. %
Columns (7)--(11) repeat the previous information for the \textit{Swift} observations. Counts  in columns (7)-(9) are reported from 0.3-10 keV, and model fluxes and luminosities in columns (10)--(11) are from 0.5-10 keV.  %
Column (12) systematic error on the X-ray flux, based on 1$\sigma$ variations in flux  from the combined \textit{Chandra} and \textit{Swift} observations on each date (see Section~\ref{sec:res:lrlx}).
}
\end{table*}
\renewcommand\arraystretch{1}

\begin{table}[htbp]
\begin{deluxetable}{l L L L}
\tablecaption{X-ray Spectral Properties \label{tab:xrayspec}}
\decimals
\tabletypesize{\small}
\tablewidth{0.9\textwidth}
\renewcommand\arraystretch{1.2}
\tablehead{
                \colhead{Date}      				&  
		\colhead{$\Gamma$}                      & 
		\colhead{$\alpha$}       			&   
		\colhead{$f_{\rm pile}$}  \\   		
		\colhead{(2015)}                            		& 
		\colhead{}		&  
		\colhead{}			& 
		\colhead{}                      
}
\colnumbers
\startdata
July 11\tablenotemark{a}    & 1.64 \pm 0.04            &  0.42 \pm 0.07                   & 0.08                    \\
July 15     & 1.75 \pm 0.07            & 0.21^{+0.69}_{-0.21}     & 0.01                    \\
July 20      & 1.71 \pm 0.06            & 0.39 \pm 0.18            & 0.05                   \\
July 23      & 1.97^{+0.08}_{-0.05}     & 0.60 \pm 0.37            & 0.03                    \\
July 23\tablenotemark{b}      & 2.04 \pm 0.04 & \nodata & \nodata \\
July 28       & 2.01^{+0.06}_{-0.02}     & 0.72^{+0.28}_{-0.41}     & 0.03                    \\
August 1    & 2.13^{+0.05}_{-0.01}     & 0.87^{+0.13}_{-0.65}     & 0.02                    \\
August 5      & 1.99 \pm 0.04            & 0.73 \pm 0.14            & 0.05                    \\
\enddata
\end{deluxetable}
\vspace{-0.8cm}
\tablenotetext{a}{Spectral parameters from a fit to the \textit{Chandra} data, freezing the column density to $\nh=8.4\times10^{21}~\cmtwo$.}
\tablenotetext{b}{Spectral parameters from a joint fit to \textit{NuSTAR}, \textit{Chandra}, and \textit{Swift} data from \datefour,  freezing the column density to $\nh=8.4\times10^{21}~\cmtwo$.}
\tablecomments{  
Column (1) observation date. %
Column (2) best-fit photon index $\Gamma$.  Unless marked otherwise, the spectral fits are from joint spectral fits to all \textit{Chandra} and \textit{Swift} data from \datetwo\ -- \dateeight, forcing a common best-fit column density across all epochs, while allowing $\Gamma$ to vary on each date.  The best-fit $\nh = 8.4 \pm 0.2 \times 10^{21} \cmtwo$. %
Column (3)  grade migration parameter from the \cite{davis01} pileup model applied to the \textit{Chandra} datasets. %
Column (4) pileup fraction in the \textit{Chandra} datasets, as calculated by the pileup model. %
All X-ray spectral fits use an absorbed power-law model {\tt phabs * powerlaw}, with abundances from \citet{anders89} and cross sections from \citet{balucinska-church92}, with updated He cross sections from \citet{yan98}.  
}
\end{table}
\renewcommand\arraystretch{1}

\subsection{Chandra Observations and Data Reduction}
\label{sec:obs:cxcdata}

 For all \textit{Chandra} observations, \src\ was placed at the aimpoint of the Advanced CCD Imaging Spectrometer \citep[ACIS;][]{garmire03} S3 chip, and the data were telemetered in FAINT mode.   To mitigate pileup, the first three observations were taken with the High Energy Transmission Grating \citep[HETG;][]{canizares05} in place to act as a filter, for which we analyzed the 0$^{th}$ order image.   The final three observations were taken without the HETG, but with the chip read in 1/8 subarray mode to reduce the exposure frame time.

The  \textit{Chandra} data reduction was performed using the \textit{Chandra} Interactive Analysis of Observations (CIAO) software v4.8  \citep{fruscione06}.  We first reprocessed the data to apply the latest calibration files, and we restricted the analysis to  the S3 chip.  We used the tool {\tt axbary} to apply barycentric corrections to all  event times, good time intervals, and aspect solutions. We next searched for time periods with elevated levels of background counts by extracting a light curve  over the entire chip (from 0.5-7~keV), after excluding \src\ and other point sources identified by the tool {\tt wavdetect}.  None of our observations displayed obvious periods of background flaring.  However, in Appendix \ref{sec:app:xray}, we describe a decision to remove the final 400 s from the first \textit{Chandra} observation (obsID 16702), for reasons related to our spectral analysis.

Finally, we extracted spectra with the tool {\tt specextract}, including response matrix files (rmf) and auxiliary response files (arf).  Although our observational setup was designed to mitigate photon pileup as much as possible, a low-level of pileup  persisted.  To apply a pileup model during spectral fitting \citep{davis01}, we extracted spectra containing all events with energies $>$0.3 keV\footnote{see \url{http://cxc.harvard.edu/ciao4.4/why/filter_energy.html}}
, and we adopted a relatively small 4 pixel radius circular extraction region centered on \src.   We applied an energy-dependent aperture correction term to the arf file to account for the small sizes of our extraction apertures.  The \textit{Chandra} X-ray properties are listed in columns (2)-(6) of  Table~\ref{tab:xray}.

\subsection{Swift Observations}
The \textit{Swift} XRT observations were taken in photon counting mode.  We analyzed the data using the {\tt HEASOFT} software, following standard procedures.  We first reprocessed each observation with the task {\tt xrtpipeline}, during which time we also created exposure maps for each observation to correct for bad columns on the detector.  We filtered the data from 0.3-10 keV, and we extracted source photons from a circular aperture with a 25 pixel radius centered at the position of \src.  We estimated the background count rate using two circular apertures (each with a radius of 25 pixels)  placed near the source, but taking care to avoid regions with enhanced soft X-ray emission from  light echoes that were observed from \src\ at this point in the decay \citep{beardmore15, beardmore16, vasilopoulos16, heinz16}.    For each observation we extracted a spectrum, we built an arf with the tool {\tt xrtmkarf} (incorporating the exposure maps created earlier), and we adopted the latest rmfs from the \textit{Swift} calibration database.  To compare to \textit{Chandra} light curves, we applied a barycentric correction to the midpoint of each \textit{Swift} snapshot (we did not obtain enough counts to extract useful \textit{Swift} light curves over shorter  timescales).   We list the X-ray properties from each XRT observation  in Table~\ref{tab:xray}. 

\begin{table*}[htbp]
\begin{deluxetable*}{l L L L L L L L L}
\tablecaption{Radio Properties \label{tab:radio}}
\decimals
\tabletypesize{\footnotesize}
\tablecolumns{9}
\tablewidth{0pt}
\renewcommand\arraystretch{1.2}
\tablehead{
                \colhead{Date}      				&  
		\colhead{$f_{5.2}$}  			&  
		\colhead{$f_{7.5}$}  			& 
		\colhead{$f_{8.6}$}   			&  
		\colhead{$f_{11.0}$}                		& 
		\colhead{$f_{8.4}$}				& 
		\colhead{$\sigma_{\rm sys}$}                 & 
		\colhead{$\left(\nu L_\nu\right)_{8.4}$}  & 
		\colhead{$\alpha_{\rm r}$}                         \\ 
		\colhead{}                            		& 
		\colhead{(mJy bm$^{-1}$)}         	& 
		\colhead{(mJy bm$^{-1}$)}         	& 
		\colhead{(mJy bm$^{-1}$)}         	& 
		\colhead{(mJy bm$^{-1}$)}         	& 
		\colhead{(mJy bm$^{-1}$)}         	& 
		\colhead{(mJy bm$^{-1}$)}         	& 
		\colhead{($10^{28}\ \ergs$)}		& 
		\colhead{}          
}
\colnumbers
\startdata
July 11\tablenotemark{a}               & 0.91 \pm 0.06     &   \nodata                       & \nodata                   & \nodata                     &   \nodata          & \nodata         & 5.2 \pm 0.7         & \nodata                \\
July 15              & 0.759 \pm 0.036      & 0.746 \pm 0.032      & 0.827 \pm 0.030      & 0.782 \pm 0.032      & 0.794 \pm 0.016      & \pm 0.046   & 4.6 \pm 0.5         & 0.07 \pm 0.08        \\
July 23              & 0.757 \pm 0.033      & 0.653 \pm 0.029      & 0.713 \pm 0.029      & 0.703 \pm 0.029      & 0.706 \pm 0.015      & \pm 0.053    & 4.1 \pm 0.5          & -0.09 \pm 0.08       \\
July 28              & 0.583 \pm 0.024      & 0.570 \pm 0.017      & 0.586 \pm 0.015      & 0.573 \pm 0.016      & 0.579 \pm 0.009      & \pm 0.057    & 3.3 \pm 0.4          & -0.02 \pm 0.07       \\
July 30              & 0.538 \pm 0.018      & 0.573 \pm 0.020      & 0.578 \pm 0.022      & 0.643 \pm 0.026      & 0.591 \pm 0.012      & \pm 0.051   & 3.4 \pm 0.4          & 0.22 \pm 0.07       \\
August 1             & 0.372 \pm 0.013      & 0.335 \pm 0.011      & 0.367 \pm 0.011      & 0.387 \pm 0.012      & 0.369 \pm 0.005     & \pm 0.057   & 2.1 \pm 0.2         & 0.07 \pm 0.05        \\
August 5             & 0.742 \pm 0.011      & 0.801 \pm 0.010      & 0.802 \pm 0.010      & 0.850 \pm 0.010      & 0.808 \pm 0.005     & \pm 0.182   & 4.6 \pm 0.5          & 0.18 \pm 0.03        \\        
\enddata
\tablenotetext{a}{VLBA observation at 5.0 GHz (see Section~\ref{sec:obs:epochone}).  We assume a flat spectral index for estimating the luminosity at 8.4 GHz.}
\tablecomments{
Column (1) observation date.
Column (2) peak flux density in the baseband centered at 5.2 GHz.
Column (3) peak flux density at 7.5 GHz.
Column (4) peak flux density at 8.6 GHz.
Column (5) peak flux density at 11.0 GHz.
Column (6) inferred flux density at 8.4 GHz from spectral fits (see Section~\ref{sec:obs:vlaspec}).   Error bars in columns (2)-(6) represent statistical uncetainties.
Column (7) systematic error on the 8.4 GHz radio flux density, based on 1$\sigma$ flux density variations within each observation (see Section~\ref{sec:res:var}).  
Column (8) radio luminosity at 8.4 GHz.  Errors  include the statistical uncertainty, and the uncertainty on the distance to the source.  
Column (9) best-fit spectral index ($f_\nu \propto \nu^{\alpha_{\rm r}}$) for the flux densities in columns 2-5. 
}
\end{deluxetable*}
\end{table*}
\renewcommand\arraystretch{1}

\subsection{\textit{NuSTAR} Observation}
\label{sec:obs:nustar}

We observed \src\ with {\em NuSTAR}  on one epoch, from 2015 July 23 UT 08:21 to
2015 July 24 UT 11:01 (ObsID 90102007011).  {\em NuSTAR} has two focal plane modules, FPMA and
FPMB, and the exposure times yielded during the observation 
were 40.2 and 39.5\,ks, respectively.  We reduced the data using HEASOFT v6.19,
NUSTARDAS v1.6.0, and the files from the 2016 July 6 calibration database (CALDB),
and we reprocessed the data to make event files using {\ttfamily nupipeline}.  We
made light curves and energy spectra using {\ttfamily nuproducts} and a circular
source extraction region with a radius of $60^{\prime\prime}$.  For background
subtraction, we used a circular region with a radius of $90^{\prime\prime}$ on the
same detector chip where the source falls.  The average source count rates in the
3--79\,keV band are $0.049\pm 0.001$ and $0.045\pm 0.001$\,s$^{-1}$ for FPMA and FPMB,
respectively.  The light curve shows that the source is near these count rates over
the duration of the observation except for the last $\sim$2\,ks of the observation
during which the count rate rose by a factor of 2--3.   In this
work, we  focus on the energy spectrum, which we rebinned with the
requirement of a signal-to-noise ratio of 5.0 in each spectral bin.

\subsection{X-ray Spectral Analysis}
\label{sec:obs:xspec}

The X-ray spectral analysis was performed with the Interactive Spectral Interpretation System v1.6.2 \citep[{\tt ISIS};][]{houck00}.  For photoelectric absorption in our fits, we used abundances from \citet{anders89} and cross sections from \citet{balucinska-church92}, with updated He cross sections from \citet{yan98}. We briefly describe  our analysis here, with more details listed in the Appendix.  For  spectra with $<$1000 counts, we binned the data to  a signal-to-noise $S/N>1.5$ per bin ($\gtrsim$2  counts); higher-count spectra were binned to $S/N > 4$ per bin ($\gtrsim$15 counts).   All fitting was performed using Cash statistics \citep{cash79}, with background counts  included in each fit.  Reported (68\%) error bars correspond to changes in the Cash statistic of $\Delta C = 1.0$ for one parameter of interest.

 On each of the six epochs we  fit an absorbed power-law model ({\tt phabs*powerlaw}) to the combined  \textit{Chandra} and \textit{Swift} observations, where we performed a joint fit by tying the column density ($\nh$) and photon index ($\Gamma$) to a common value, but allowing the normalizations of each dataset to independently vary.    For the \textit{Chandra} datasets, we used the \citet{davis01} pileup model to correct for mild effects of photon pileup (the \textit{Swift} data did not suffer from any pileup).   Within the \citet{davis01} model, we fixed the {\tt psfrac} parameter to 0.95 (the fraction of the incident energy that falls on the central 3x3 pixels), and we left the `grade migration parameter' $\alpha$ free to vary (the probability of retaining $n$ `piled' events as a single event is $p \sim \alpha^{n-1}$).   The inclusion of the \textit{Swift} data increased the number of counts, especially at soft X-rays for the first three epochs when the \textit{Chandra} HETG was in place, and it also assisted in constraining the pileup correction (see Appendix~\ref{sec:app:xray} for further discussion).    
 
 The best-fit spectral parameters are presented in Figure~\ref{fig:confmap} in the Appendix, along with a sample spectral fit to our \textit{Chandra} observation from \dateeight\ (our highest $S/N$ spectrum) in Figure~\ref{fig:chspec}.   The best-fit $\nh$ values on each epoch were consistent with each other within the errors.  Therefore, we performed another joint fit where we forced a common $\nh$ across all six epochs (but we allowed $\Gamma$ to vary on each epoch).  We found a best-fit $\nh=8.4 \pm 0.2 \times 10^{21}\ \cmtwo$, and the best-fit photon indices are presented   in Table~\ref{tab:xrayspec}, which are the values adopted throughout the rest of the text (except on \datefour; see next paragraph).     The level of pileup in the \textit{Chandra} observations from \datetwo--\dateeight\ was mild, from 2-5\%. 
 
 On \datefour\ we also obtained a simultaneous observation with \textit{NuSTAR}.  To  improve the spectral model, we fit the \textit{Chandra}, \textit{Swift}, and \textit{NuSTAR} data from \datefour\ (including both the \textit{NuSTAR} FPMA and FPMB spectra), freezing the column density to $8.4 \times 10^{21}~\cmtwo$, and forcing a common $\Gamma$.  The best-fit photon index $\Gamma=2.04\pm 0.04$ was consistent with the best-fit value when only considering the \textit{Chandra} and \textit{Swift} data ($\Gamma=1.97^{+0.08}_{-0.05}$), thereby indicating that our spectral results can be extended toward higher energies.  Throughout the remainder of the text, we adopt $\Gamma=2.04\pm0.04$ on \datefour.

\subsection{VLA Radio Observations}
\label{sec:obs:vla}

 A total of six VLA observations (project code SG0196) were taken   between 2015 July 15 -- 2015 August 5,  with on-source exposure times ranging from 8 -- 93 min (see Table~\ref{tab:obslog}).   Each VLA observation was scheduled to obtain as much strictly simultaneous coverage with \textit{Chandra}  as possible.  As noted earlier, we did not obtain VLA observations on our second \textit{Chandra} epoch (July 20), but we did obtain an extra VLA observation on July 30 (for which there was no corresponding X-ray observation).

The VLA was in its most extended A configuration, with a maximum baseline of 30 km.  We made use of the VLA `subarray' mode, where approximately half of the VLA antennas observed at 4--8 GHz, and the other half at 8--12 GHz.    We separated the two 1024-MHz basebands within each observing band to provide the broadest possible spectral coverage, while avoiding known radio frequency interference.  Each 1024-MHz baseband comprised eight spectral windows, each made up of 64 2-MHz channels.  The central frequencies of the basebands were 5.2, 7.5, 8.6, and 11.0 GHz.  The subarrays provided a valuable frequency lever arm for investigating the radio spectrum. 

The radio analysis was performed using standard procedures within the Common Astronomy Software Application v4.5 \citep[{\tt CASA};][]{mcmullin07}.  We calibrated each 1024-MHz baseband separately, using the \citet{perley13} coefficients within the {\tt setjy} task to set the amplitude scale.   We selected our amplitude calibrator according to the local sidereal time of each observation, using 3C 286 on July 15 and Aug 5, and 3C 48 on all other epochs.     At all epochs, we solved for the complex gain solutions toward \src\ by using the secondary calibrator source J2025+3343.  On July 28, we did not obtain any usable scans of a primary flux calibrator.  So, we  manually set the amplitude scale in {\tt setjy} to the flux density of J2025+3343, which was determined by interpolating the flux density of J2025+3343 from the two surrounding epochs (July 23 and 30) to July 28 (the flux densities on July 23 and 30 were calculated  by the task {\tt fluxscale} when bootstrapping the amplitude gain solutions to J2025+3343 on those epochs).   Over our three-week campaign, we measured flux density variations for our phase calibrator J2025+3343 at the 3, 2, 1, and 1\% levels  (1$\sigma$)  at 5.2, 7.5, 8.6 and 11.0 GHz, respectively.  We included corresponding systematic uncertainties on flux densities from July 28.

We next imaged the field surrounding \src\ with the task {\tt clean}, using Briggs weighting with a robust value of 1, and two Taylor terms to model the frequency dependence of sources in the field.  We placed  outlier fields on two bright sources within the primary beam, so that their sidelobes did not influence the final \src\ image.  We achieved 1$\sigma_{\rm rms}$ sensitivities from $\approx$0.010--0.035 mJy bm$^{-1}$, depending on the exposure time and frequency.  These sensitivities are consistent with the theoretical noise limit of the VLA (for 13 antennas per frequency).  Finally, we measured the flux density of \src\ at each epoch by fitting a point source in the image plane using the task {\tt IMFIT} (see Table~\ref{tab:radio}).

\subsubsection{Radio Spectral Indices and 8.4 GHz Flux Densities}
\label{sec:obs:vlaspec}

For each radio epoch, we measured the radio spectral index $\alpha_{\rm r}$ ($f_\nu \propto \nu^{\alpha_{\rm r}}$)  by fitting a power-law to the four flux density measurements  (via a weighted least squares fit).    We estimated the uncertainty on the spectral index, $\sigma_{\alpha_{\rm r}}$, through Monte Carlo simulations.    We added simulated statistical noise to each flux density (based on a Gaussian distribution with a standard deviation set to the uncertainty on each flux density measurement), and we also randomly adjusted the central frequency of each flux density  across each 1024 MHz baseband (assuming a uniform distribution in frequency).  We then fit a power-law to  each dataset with random noise added, and we repeated 1000 times for each epoch.  As expected, the distribution of 1000 $\alpha_{\rm r}$ measures on each epoch followed an approximately Gaussian distribution centered about zero.  For $\sigma_{\alpha_{\rm r}}$, we adopted the standard deviation on 1000 simulated $\alpha_{\rm r}$ measures (which we confirmed is comparable to the 68\% confidence interval).   We used the above spectral fits to calculate radio flux densities (and errors) at 8.4 GHz, which are the values we generally adopt throughout this paper when displaying radio fluxes and luminosities in figures, in order to ease comparisons to the literature.

\subsection{Other Multiwavelength Observations on \dateone}
\label{sec:obs:epochone}

We also considered a 39 ks \textit{Chandra} observation granted through Director's Discretionary Time (DDT) that was taken on  2015 July 11 (obsID 17701; PIs Neilsen and Altamirano), four days before our first ToO observation.   This DDT observation used the HETG, and it was obtained primarily to study a disk wind through high-resolution spectroscopy.  Here, we considered only the 0$^{th}$ order grating image, in order to extend our time coverage during the decay into quiescence.  Analysis on this observation was performed identically to our six ToO observations.   We performed two spectral fits (of the form {\tt phabs*powerlaw}), where we first allowed  $\nh$ to vary as a free parameter (yielding $\nh= 1.0 \pm 0.1 \times 10^{22}~\cmtwo$, $\Gamma=1.79 \pm 0.10$; see Figure~\ref{fig:confmap}),  and then we fixed $\nh$ to $8.4 \times 10^{21} \cmtwo$ (i.e., the best-fit value from Section~\ref{sec:obs:xspec}).  The latter fit yielded $\Gamma=1.64\pm0.04$, which is the value we adopt throughout the text (see Table~\ref{tab:xrayspec}).  We  applied the \citet{davis01} pileup model  during these fits, and we found  8\% pileup. 

No VLA observations were taken on \dateone.  However, there was a Very Long Baseline Array (VLBA) radio program (PI Miller-Jones; project code BM421), from which we extracted radio information on \dateone.  Although no VLBA observations were taken simultaneously with \textit{Chandra}, a 25 min portion of a VLBA observation from UT 07:03-07:28 ($f_\nu = 0.91 \pm 0.06$ mJy at 5GHz) was obtained simultaneously with a \textit{Swift} snapshot ($f_{\rm 0.5-10~keV} = 1.98 \pm 0.12 \times 10^{-11}~\flux$; model flux calculated in {\tt ISIS} assuming  $\nh=8.4\times10^{21}~\cmtwo$ and $\Gamma=1.64$).  For placing these data on radio/X-ray luminosity correlations,  we  adopted   $\lr= \left( 5.2 \pm 0.7\right) \times 10^{28}\  \ergs$ at 8.4 GHz (assuming a flat radio spectrum) and   $L_{\rm 1-10~keV} = \left(1.16 \pm 0.15\right) \times 10^{34}~\ergs$.   Details on the VLBA reduction will be provided in an upcoming publication (Miller-Jones et al.\ in prep).

\section{Results}
\label{sec:res}

\subsection{Long-Term Flux and Spectral Evolution}
\label{sec:res:longterm}

\begin{figure}
\includegraphics[scale=0.58]{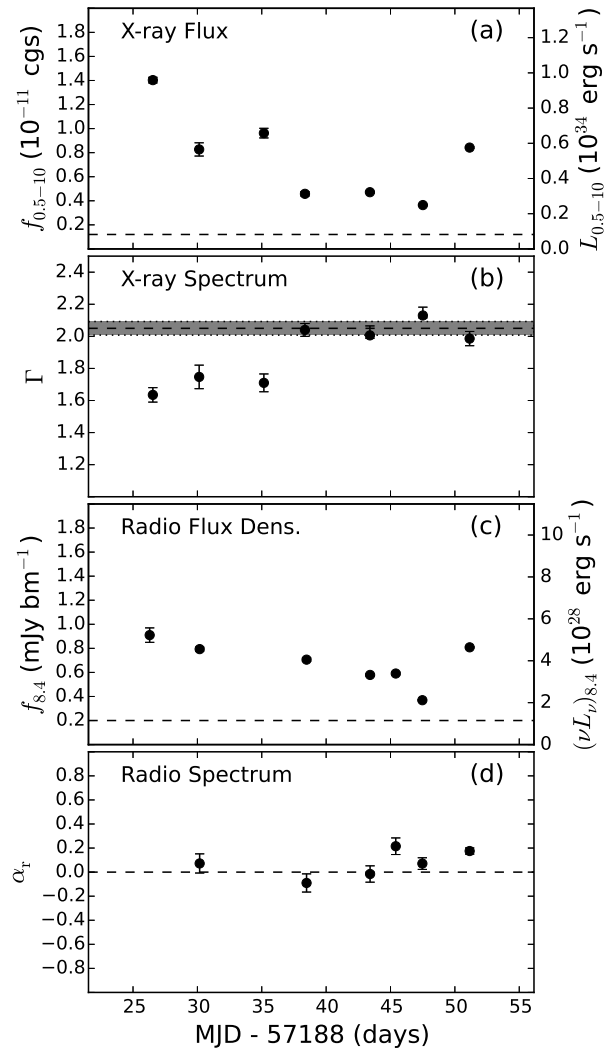}
\caption{(a) \textit{Chandra} X-ray light curve during  our three-week  campaign.  Unabsorbed X-ray fluxes (luminosities) from 0.5-10 keV are labeled on the left (right) vertical axis.  The dashed line illustrates the  average pre-outburst X-ray flux \citep{bernardini14} (b) Evolution of the X-ray photon index.  The dashed line and gray shaded region denote the pre-outburst $\Gamma$ and 1$\sigma$ uncertainty from \citet{reynolds14}.  (c) Interpolated radio flux density light curve at 8.4 GHz (see left vertical axis); corresponding luminosities at 8.4 GHz are labeled on the right vertical axis, and the dashed line shows the pre-outburst radio flux density \citep{corbel08}.  (d) Radio spectral index $\alpha_{\rm r}$ (we do not have radio spectral constraints from the \dateone\ VLBA observation).  The dashed line marks a flat radio spectrum for reference.   In all panels, the time axis is referenced to the discovery date of the outburst (2015 June 15).}
\label{fig:decay}
\end{figure}

In Figure~\ref{fig:decay} we show the X-ray and radio  light curves during our three-week campaign, along with the corresponding evolution of the X-ray photon index $\Gamma$ and the radio spectral index $\alpha_{\rm r}$.  Throughout our campaign, the average \textit{Chandra} X-ray fluxes (0.5-10 keV) are  a factor of 3-12 brighter than the average pre-outburst X-ray flux of \src.  Although there is an overall trend of decreasing flux with time,  there is also superposed variability, so that the decay is not  monotonic.   Figure~\ref{fig:decay}b displays a clear X-ray spectral softening, where the spectrum is relatively hard toward the beginning  ($\Gamma=1.64\pm 0.04$  on \dateone) and settles near the pre-outburst value of $\Gamma \approx 2$ by the end of our campaign.\footnote{Note that we would still observe an X-ray spectral softening  if we were to adopt $\Gamma=1.79\pm0.10$ on \dateone\ when allowing the column density to vary as a free parameter.}  
Even though \src\ re-brightens on our final epoch (\dateeight) to a flux comparable to \datetwo\ ($\approx8\times 10^{-12}~\flux$), its X-ray spectrum remains soft on \dateeight\ ($\Gamma=1.99\pm0.04$) compared to on \datetwo\ ($\Gamma=1.75\pm0.07$).  
 
The radio spectrum is consistent with being flat/inverted throughout our entire campaign, and only on \datesix\ and \dateeight\ does it appear to be inverted at a meaningful level ($>$$3\sigma$) (Figure~\ref{fig:decay}d).  Throughout, we adopt  0.2 mJy ($1.1 \times 10^{28}~\ergs$ at 8.4 GHz; \citealt{corbel08}) as the pre-outburst  radio flux density, so that the radio emission is a factor of $\approx$2-5 brighter during our campaign compared to pre-outburst.  However,  we note that both \citet{miller-jones08} and \citet{rana16} observe slightly higher average radio flux densities of 0.3 mJy in quiescence, implying  that \src\ may have temporarily reached its pre-outburst radio flux level during our campaign on \dateseven.

\begin{figure}
\begin{center}
\includegraphics[scale=0.58]{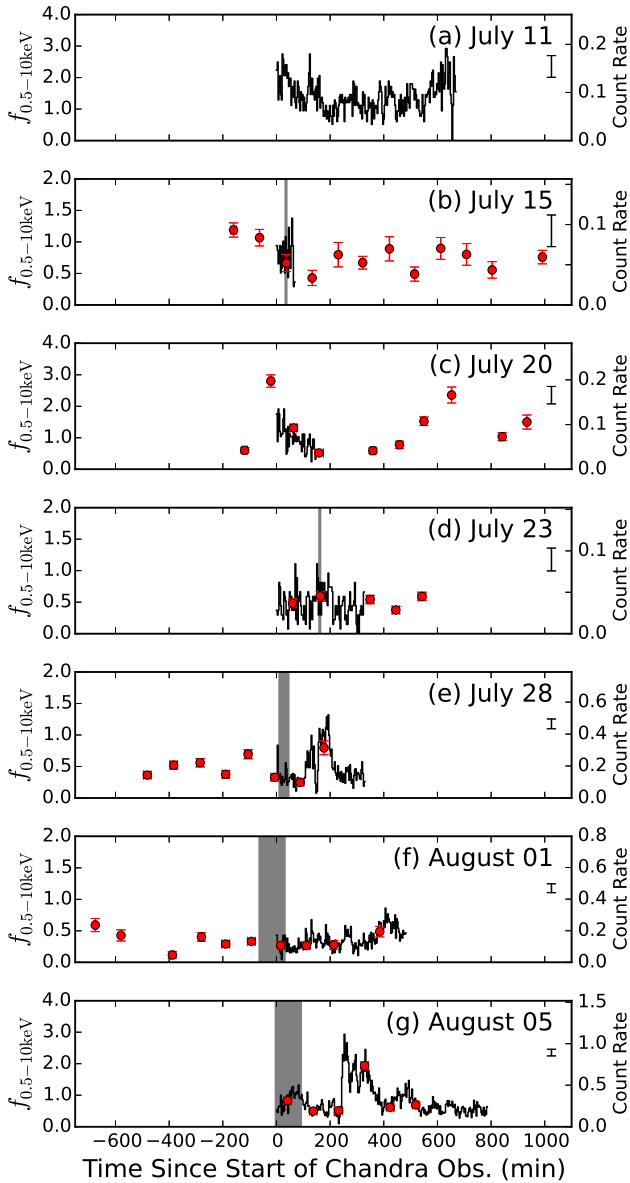}
\caption{X-ray Light curves of seven Chandra observations binned by 3 min (black solid lines), with typical error bars illustrated in the top right of each panel.  X-ray fluxes from $\approx$1-2 ks \textit{Swift} snapshots are overplotted (red circles).  Shaded regions mark when the VLA observed.  X-ray fluxes are from 0.5-10 keV, in units of 10$^{-11}\ \flux$ (see left vertical axes), and count rates for the \textit{Chandra} observations are listed on the right vertical axes..  The x-axis is referenced to the start time of each \textit{Chandra} observation (see Table~\ref{tab:obslog}; times from all telescopes are barycentered). }
\label{fig:xlc}
\end{center}
\end{figure}

\subsection{Intraday Variability}
\label{sec:res:var}

For each epoch, we create X-ray  and radio light curves to explore variability on timescales of minutes to hours.   X-ray light curves over 3 min time bins are displayed in Figure~\ref{fig:xlc} for all seven \textit{Chandra} observations, with fluxes  from  \textit{Swift} snapshots overplotted.   Five of the \textit{Chandra} observations contain time periods with strictly simultaneous VLA coverage (ranging from 8-93 min), which we highlight as gray regions in Figure~\ref{fig:xlc}.  For each light curve, we quantify the level of variability with the fractional rms variability amplitude statistic $\fvar$ \citep[see, e.g.,][]{vaughan03}, which we report in Table~\ref{tab:fvar}, incorporating all \textit{Chandra} and \textit{Swift} observations.  The $\fvar$ statistic probes timescales as short as 3 min from each \textit{Chandra} observation, and timescales as long as 8-19 hours (depending on the date), which corresponds to the range of times between the first and final \textit{Swift} snapshots on each epoch (see Table~\ref{tab:obslog}).  All seven X-ray light curves display variability at the $\fvar \approx 20-55$\% level.  Throughout, we (somewhat arbitrarily) require $\fvar/\sigma_{\fvar}>1$ to claim  variability, and we refer to observations displaying $1 \leq \fvar/\sigma_{\fvar} < 3$ as mildly variable, where $\sigma_{\fvar}$ is the  statistical uncertainty on $\fvar$.   Photon pileup in our \textit{Chandra} observations could suppress the observed X-ray variability so that our $\fvar$ values are underestimated by up to a factor of 0.84 during the most extreme flares (e.g., the flare from 200-400 min into the \dateeight\ observation; see Appendix~\ref{sec:app:xray}).  However, during the bulk of our observations, effects from pileup are generally more mild.   We do not obtain a sufficient number of counts in each 3 min bin to investigate short-term spectral variability in the X-ray, although we note in Appendix~\ref{sec:app:xray} that we do not see evidence for X-ray spectral variability within individual \textit{Chandra} exposures binned by count rate.    For completeness, we search for quasiperiodic oscillations in our highest count rate \textit{Chandra} observation (\dateeight), and we see no evidence from 0.001 - 1 Hz.   

\begin{table*}[htbp]
\begin{deluxetable*}{l C C C C c c }
\tablecaption{Variability \label{tab:fvar}}
\tablewidth{\textwidth}
\tablecolumns{11}
\renewcommand\arraystretch{1.2}
\tablehead{
		 &    
		\multicolumn{2}{c}{Full Exposures}   &  
		\multicolumn{3}{c}{Strictly Simultaneous} & 
		      \\  
                \colhead{Date}      				&  
		\colhead{$F_{\rm var, xray}$}  		&  
		\colhead{$F_{\rm var, radio}$} 		& 
		\colhead{$F_{\rm var, xray}$}              & 
		\colhead{$F_{\rm var, radio}$}		& 
		\colhead{$t_{\rm exp, sim}$}            & 
		\colhead{$p$}	 \\ 
		\colhead{(2015)}   & 
		\colhead{}   & 
		\colhead{}   & 
		\colhead{}   & 
		\colhead{}   & 
		\colhead{(min)}   & 
		\colhead{}    
}		 
\colnumbers
\startdata
Jul 11          &  0.25 \pm  0.03      & \nodata              & \nodata              & \nodata              & \nodata              & \nodata            \\
Jul 15          &  0.18 \pm  0.10      &  0.04 \pm  0.19       &  0.11 \pm  0.85      &  0.04 \pm  0.19         & 8              & \nodata         \\
Jul 20          &  0.39 \pm  0.05      & \nodata                & \nodata              & \nodata              & \nodata                 & \nodata         \\
Jul 23          &  0.24 \pm  0.09      &  0.06 \pm  0.17       &  0.37 \pm  0.07      &  0.06\pm  0.17              & 8              & \nodata         \\
Jul 28          &  0.54 \pm  0.05      &  0.09 \pm  0.04         &  0.29 \pm  0.13      &  0.09 \pm  0.04            & 35              & 0.2           \\
Jul 30          & \nodata              &  0.04 \pm  0.08      & \nodata              & \nodata              & \nodata                  & \nodata         \\
Aug 01          &  0.36 \pm  0.05      &  0.12 \pm  0.03          &  0.39 \pm  0.13      &  0.09\pm  0.07            & 32              & 0.4           \\
Aug 05          &  0.55 \pm  0.02      &  0.22 \pm  0.01          &  0.28 \pm  0.03      &  0.22 \pm  0.01                & 93              & 0.003           \\
 \enddata
 \end{deluxetable*}
 \vspace{-0.8cm}
\tablecomments{\footnotesize  Column (1) observation date.  
Column (2) the $F_{\rm var}$ statistic (with 1$\sigma$ statistical uncertainties), which quantifies the rms flux variability for the full combined \textit{Chandra} and \textit{Swift} X-ray light curves (\dateone\ is based only on \textit{Chandra} observations).  
Column (3) $F_{\rm var}$ for the full duration of each VLA radio light curve.  
Columns (4)-(5) same as previous two columns, but over the time periods with strict simultaneity between \textit{Chandra} and the VLA (\textit{Swift} data omitted).  Note that columns (3) and (5) only differ on  \dateseven, which is the only time that a VLA observation began before \textit{Chandra}.
Column (6) duration of  simultaneous overlap (reported as the total VLA time on source).  
Column (7)  the probability of no correlation between the radio and X-ray fluxes, for the three epochs with the most simultaneous overlap.
}
\end{table*}
\renewcommand\arraystretch{1}

\begin{figure}
\begin{center}
\includegraphics[scale=0.55]{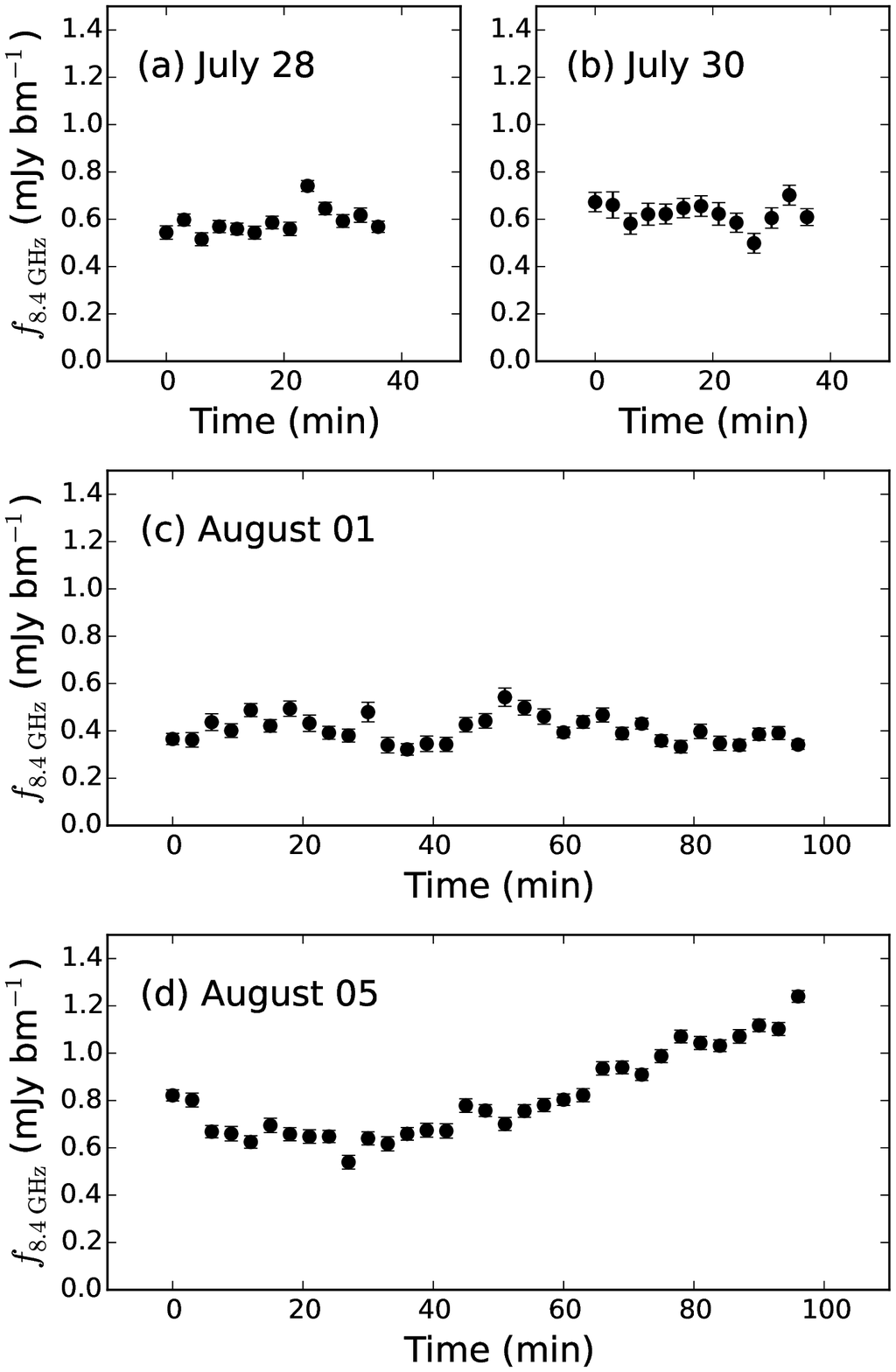}
\caption{Radio light curves for the four epochs with the longest amount of radio coverage, binned by 3 min.  The x-axis is  referenced to the start time of each  VLA observation.}
\label{fig:rlc}
\end{center}
\end{figure}

\begin{figure}
\begin{center}
\includegraphics[scale=0.55]{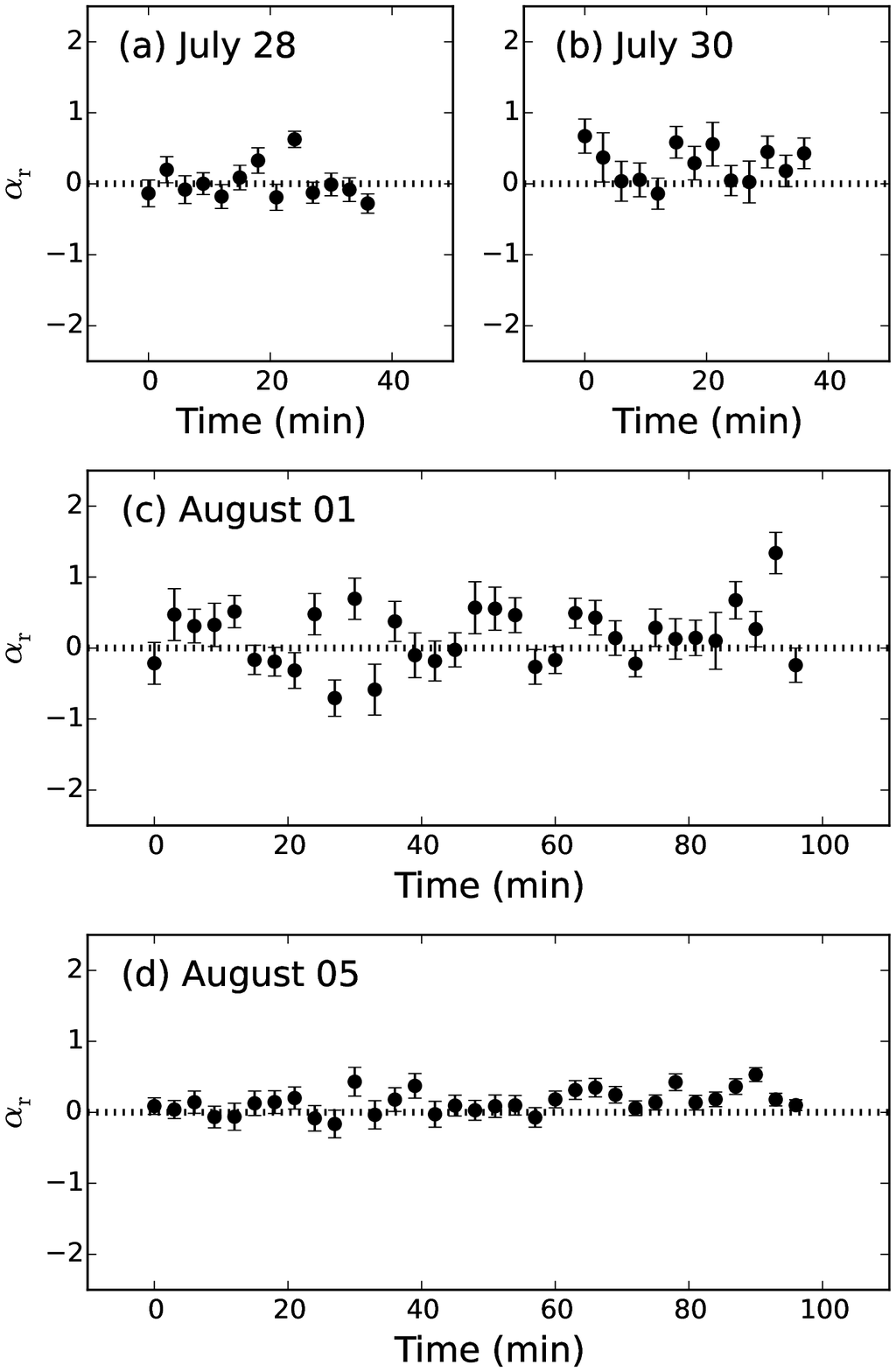}
\caption{Variation of radio spectral index with time for  the four longest radio epochs (binned by 3 min).   The dotted lines mark $\alpha_{\rm r}=0$  for reference.  The x-axis is referenced to the beginning of each VLA observation.}
\label{fig:rspindlc}
\end{center}
\end{figure}

For our six VLA epochs, we split each radio observation into 3 min time bins, which we  image  at each central frequency (following the same procedure described in Section~\ref{sec:obs:vla}).  We have sufficient radio signal to also investigate radio spectral variability.  For each time bin, we fit a power-law to the radio spectrum to measure $\alpha_{\rm r}$ and  infer the radio flux density at 8.4 GHz, following the procedure in Section \ref{sec:obs:vlaspec}.  For each time bin, we typically measure flux densities to accuracies of $\sigma_{f} \approx \pm 0.03-0.04$ mJy bm$^{-1}$   and spectral indices to $\sigma_{\alpha_{\rm r}} \approx \pm 0.1-0.3$.    Light curves for the 8.4 GHz radio flux density and for the radio spectral index $\alpha_{\rm r}$ are displayed in Figures~\ref{fig:rlc} and \ref{fig:rspindlc}, respectively (we omit our first two radio epochs because they  contain $<$10 min on source).     \src\ is displays strong radio variability during our  final \dateeight\ epoch ($\fvar=0.22\pm 0.01$), and to lesser extents on \dateseven\  ($\fvar = 0.12\pm0.03$) and \datefive\ ($\fvar=0.09 \pm 0.04$; see Table~\ref{tab:fvar}). Intriguingly, we do not see rapid intraday fluctuations in $\alpha_{\rm R}$, in contrast to \citet{rana16} who observe the radio spectrum to fluctuate between optically thin and optically thick over 10 min time intervals (although we note that we observed when \src\ was up to a factor of three radio brighter, and a detailed comparison is out of the scope of this paper).

\begin{figure*}
\begin{center}
\includegraphics[scale=0.68]{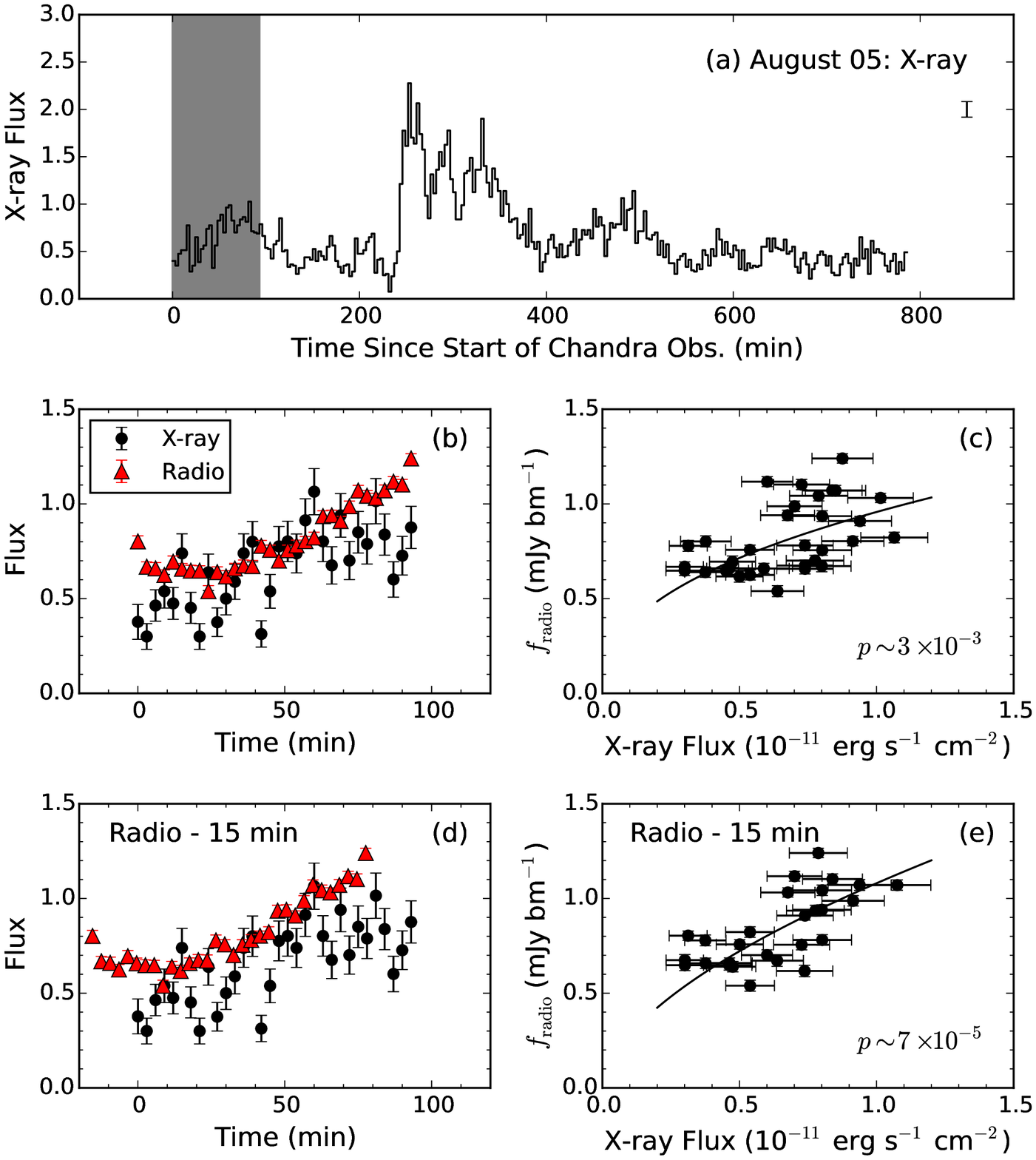}
\caption{(a) The hard (1-10 keV) \textit{Chandra} X-ray light curve from \dateeight\ (binned by 3 min), with the  shaded region denoting the time period with strictly simultaneous VLA observations.   A typical error bar is illustrated in the top right corner.  (b) X-ray (black circles) and radio (red triangles) light curves during the period of strict simultaneity.   (c) Radio vs.\ X-ray emission.  A Pearson correlation test indicates a (marginal) positive correlation at the $p \approx 3\times10^{-3}$ level.  The black solid shown shows the best-fit radio/X-ray correlation during the 93 min of strict simultaneity ($\lr \propto \lx^{0.4 \pm 0.2}$).  (d) Same as panel (b), except with the radio data delayed by 15 min.  (e) Same as panel (c), except with X-ray light curves re-extracted after factoring in the 15 min radio delay.   After correcting for the time delay, a Pearson correlation test indicates a stronger correlation ($p \approx 7 \times 10^{-5}$), and $\lr \propto \lx^{0.6\pm0.2}$.  All X-ray fluxes are from 1-10 keV (in units of 10$^{-11}~\flux$), and all radio flux densities are at 8.4 GHz (in units of mJy bm$^{-1}$);  the normalizations of the radio and X-ray light curves in panels (b) and (d) correspond to the aforementioned units.}
\label{fig:aug05lc}
\end{center}
\end{figure*}

Finally, we  consider the five  epochs where the \textit{Chandra} and VLA observations contain periods of strict simultaneity (\datetwo, \datefour, \datefive, \dateseven, and \dateeight).  For a proper comparison, we apply barycentric corrections to the times of each 3 min VLA time bin, and we extract \textit{Chandra} light curves over strictly simultaneous 3 min bins.   The $\fvar$ statistic is reported in Table~\ref{tab:fvar} for each portion of these five radio and X-ray observations with strict simultaneity (we also report in Table~\ref{tab:fvar} the length of strict overlap).   We detect X-ray variability in  4/5 epochs (although most significantly on \dateeight), and we detect radio variability only on \dateeight.  For the final three epochs with $>$30 min of strict simultaneity (\datefive, \dateseven, and \dateeight), we search for correlated X-ray and radio variations using a Pearson correlation test.  There is a hint for a weak, but not highly significant, correlation on \dateeight\ ($p=0.003$ that no correlation is present; see Figure~\ref{fig:aug05lc} for a comparison of the radio and X-ray light curves on \dateeight), which we describe further in the next subsection.  There is no evidence for correlated X-ray/radio variations on either \datefive\ ($p=0.2$) or \dateseven\ ($p=0.4$).  

\subsubsection{August 5: Correlated Variability and a Radio Lag?}
\label{sec:res:tdelay}

The \dateeight\ epoch shows strong variability in both the X-ray and radio, with a marginally significant correlation between the X-ray and radio fluxes.  Examination of the X-ray and radio light curves (Figures~\ref{fig:xlc}, \ref{fig:rlc}, and \ref{fig:aug05lc}) suggests that \src\  began a small flare at the time when both \textit{Chandra} and the VLA were observing.    We calculate the cross-correlation function (CCF) for the 1-10 keV X-ray and 8.4 GHz radio light curves\footnote{CCF results on the light curves from \datefive\ and \dateseven\ are inconclusive, which is to be expected since we do not observe  obvious X-ray flares during the (shorter) $\approx30$ min periods of overlap on those dates.}  
 over the 93 min period of overlap (using 3 min time bins that are strictly simultaneous; we adopt 1-10 keV X-ray fluxes here for consistency with our radio/X-ray luminosity correlation analysis in  Section~\ref{sec:res:lrlx}, but results are unchanged if we adopt 0.5-10 keV fluxes).  Our radio and X-ray light curves lack sufficient time coverage to define a non-flaring continuum level.  Therefore, when calculating the CCF, we apply the locally normalized discrete correlation function algorithm described by \citet{lehar92}.  This algorithm is similar to the discrete correlation function \citep[e.g.,][]{edelson88}, except, at a given time delay,  the first and second moments of each time series are calculated by considering only the subset of data pairs within each time delay bin, instead of over the entire time series.

\begin{figure}
\begin{center}
\includegraphics[scale=0.5]{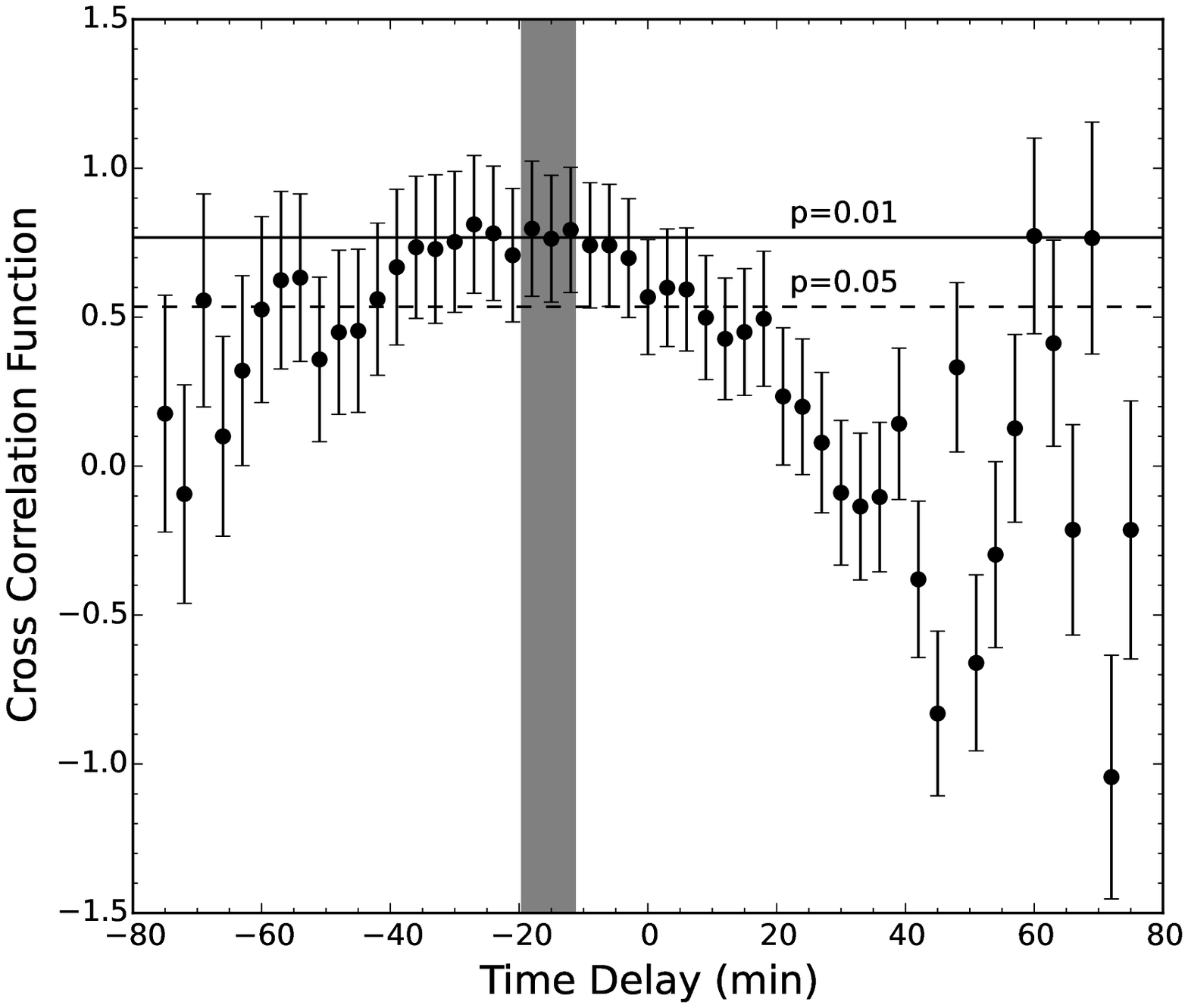}
\caption{Cross correlation function (CCF) for the strictly simultaneous X-ray and radio light curves on \dateeight.  Negative time delays mean that the radio emission lags the X-ray emission.  The CCF shows marginal evidence for the radio emission lagging the X-ray by $15 \pm 4$ min (the shaded region illustrates the $\pm$1$\sigma$ confidence interval on the time delay).  The solid and dashed horizontal lines mark the $p=0.01$ and $p=0.05$ probabilities, respectively, that the CCF peak is due to random fluctuations and/or uncorrelated variability (see Section~\ref{sec:res:tdelay}).}
\label{fig:ccf}
\end{center}
\end{figure}

The CCF over these 93 min of strict simultaneity is displayed in Figure~\ref{fig:ccf}.  Error bars are calculated through the following bootstrapping method that simulates the expected CCF for uncorrelated variations:  we  randomize each radio and X-ray light curve 1000 times (after adding random statistical noise to each data point according to the flux measurement uncertainties), and we then calculate the CCF on each realization of the data.  The error bars in Figure~\ref{fig:ccf} display the standard deviations at each time delay for the 1000 bootstrapped CCFs.  The CCF peaks (i.e., shows the strongest positive correlation) at a time delay of $\Delta t = -\tdelay \pm \tdelayErr$ min,  where a negative time delay indicates that the radio emission lags the X-ray emission.\footnote{To estimate the time delay ($\Delta t$) and error ($\sigma_{\Delta t}$), we consider data points in the CCF between $-48 < \Delta t < 18$ min (which corresponds to the full-width half-max of the peak in the CCF).  We then calculate $\Delta t$  and $\sigma_{\Delta t}$ respectively as the average time  and standard deviation, weighted by the CCF.}  
{The value of the CCF at $\Delta t = -15$ min is $0.76 \pm 0.22$.  To estimate the statistical significance of delayed correlated variability, we use the above simulations to calculate a global significance level (following \citealt{bell11}).   We determine the fraction of simulated CCF values at any time delay with a value $>$0.76, and we find $p=0.01$.  This global significance level accounts for stochastic fluctuations as well as false detections from any intrinsic yet uncorrelated variability within each light curve (see \citealt{bell11} for details).

As noted earlier, the observed radio and X-ray emission are correlated at a marginally significant level ($p=0.003$, from a Pearson correlation test).  However, if we remove the radio lag by shifting the radio light curve by 15 min (Figure~\ref{fig:aug05lc}d) and then re-extract the X-ray light curves over 3 min bins, then the correlated radio and X-ray variability becomes more statistically significant ({$p = 7 \times 10^{-5}$;  Figure~\ref{fig:aug05lc}e).  For completeness, we also perform a linear regression to the radio/X-ray correlations over the 93 min of strict simultaneity, and we find a marginally steeper slope after removing the radio lag ($f_r \propto f_x^{0.42\pm0.15}$ as observed, and $f_r \propto f_x^{0.59\pm0.17}$ after removing the radio delay; see Section \ref{sec:res:lrlx} for a description of our fitting method). 

   We stress that we consider the evidence for a radio time delay to be tentative at the moment, as the $p=0.01$ chance of a random correlation yields only a marginal detection.  Furthermore, the light curves supporting this CCF analysis are not optimal, as we did not observe the beginning of the flare in the X-ray, nor did we observe the maximum of the flare in the radio.  Nevertheless, we report the CCF results here in order to highlight a  result that merits further investigation, but we proceed cautiously with our interpretation (see Section~\ref{sec:disc:jetsize}).

\subsection{Radio -- X-ray Luminosity Correlation} 
\label{sec:res:lrlx}

In Figure~\ref{fig:lrlxfull} we add our five new epochs of simultaneous VLA/\textit{Chandra} observations, along with the  VLBA/\textit{Swift} observations from \dateone,  to the radio/X-ray luminosity correlation for \src.  We  compare to   quasi-simultaneous data  from  the 1989 outburst and to  two epochs of simultaneous observations in quiescence,  as described in Section~\ref{sec:obs:lit}.   Our 2015 campaign filled in a luminosity regime that was not well covered during the 1989 outburst, and to our knowledge, no other radio telescope covered these luminosities in 2015. 

The error bars displayed in Figure~\ref{fig:lrlxfull} for our 2015 campaign include statistical uncertainties from the flux measurements (with the distance uncertainty propagated, to ease comparisons to the literature), and a systematic error set to the standard deviations of the intraday flux variability on each epoch (see Tables~\ref{tab:xray} and \ref{tab:radio}).   For the July 11 VLBA observation, we assume a systematic error of 0.15 dex.  Errors for the data points taken from the literature are described in Section \ref{sec:obs:lit} (where we assume that flux uncertainties for the quasi-simultaneous observations are dominated by variability induced systematics).   Following \citet{gallo14}, we  fit a  function of the form $\left( \log \lr - 29\right) = b + m \left( \log \lx - 35 \right)$ to the updated radio/X-ray luminosity correlation, using the Bayesian linear regression  technique of \citet{kelly07}.  We obtain nearly identical results as \citet{corbel08} and \citet{gallo14}, who consider  the 1989 and 2003 data.  We find $b = 0.40 \pm0.04$, $m=0.54 \pm 0.03$, and $\sigma_{\rm int} = \pm 0.06 \pm 0.03$.

\begin{figure*}
\begin{center}
\includegraphics[scale=0.7]{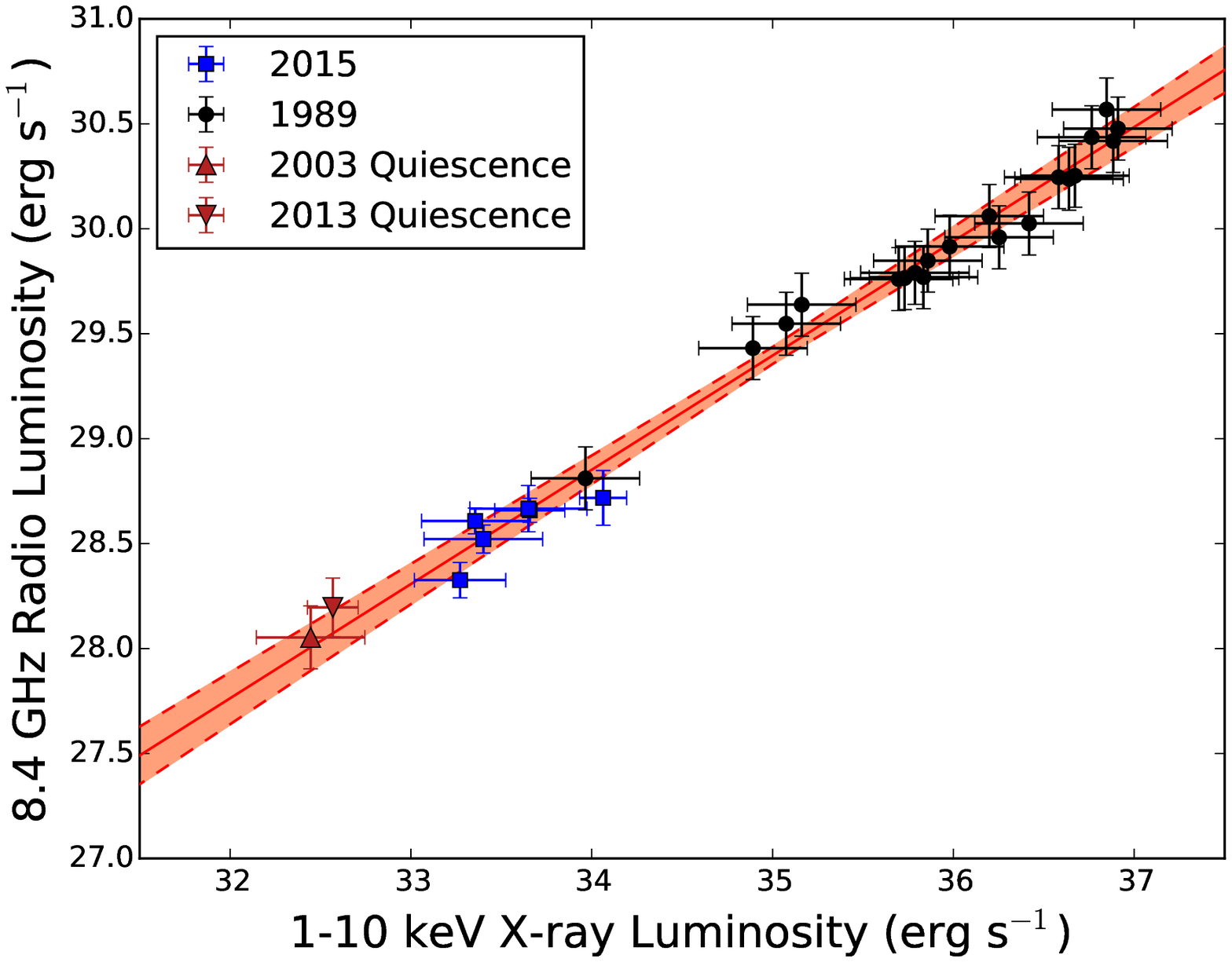}
\caption{The radio/X-ray luminosity correlation for \src.  Blue squares show observations from our 2015 campaign, including the simultaneous VLBA and \textit{Swift} observations on \dateone.  Black circles show data from the 1989 decay \citep{corbel08}, and the red triangles show two pre-outburst epochs: 2003 \citep{corbel08, hynes09} and 2013 \citep{rana16}.  The red solid line  and shaded region shows the best-fit to the radio/X-ray correlation ($\lr \propto \lx^{0.54}$) and the 1$\sigma$ uncertainty.}  
\label{fig:lrlxfull}
\end{center}
\end{figure*}

Since all of our 2015 X-ray observations are longer than our radio observations, we also  re-image our radio and X-ray observations to only include strictly simultaneous time periods, in order to investigate how non-simultaneity may influence the radio/X-ray correlation.  We find the difference to be negligible, and well within the attributed measurement errors and scatter about the luminosity correlation: the radio luminosities are hardly affected, and no X-ray luminosity changes by more than 0.2 dex.  We similarly do not find any impact to the radio/X-ray correlation if we  consider periods of strict overlap after removing the 15 min radio time delay.

\subsection{Comparison Observations from the Literature}
\label{sec:obs:lit}
Here, we describe comparison data from other X-ray and radio campaigns on  \src\ in the literature.  Our comparison data is  comprised  primarily of quasi-simultaneous radio and X-ray observations from the 1989 outburst, as compiled by \citet{corbel08}, and two epochs of simultaneous radio and X-ray observations in quiescence (pre-2015 outburst), one from the VLA/\textit{Chandra} in 2003 \citep{hynes04, hynes09, corbel08}, and one from the VLA/\textit{NuSTAR} in 2013 \citep{rana16}.  We also adopt X-ray spectral parameters from the literature based on three additional X-ray observations taken pre-outburst from \textit{Chandra}, \textit{XMM-Newton}, and \textit{Suzaku} \citep[][]{reynolds14} and one \textit{Chandra} observation taken after the outburst in 2015 November (\citealt{tomsick15}; see Section~\ref{sec:obs:lit:spec}). 

\subsubsection{\citet{corbel08}}
\label{sec:obs:lit:c08}

\citet{corbel08}  assemble a total of 20 epochs of quasi-simultaneous radio and X-ray observations of \src\ during the decay of its 1989 outburst, with luminosities ranging from $10^{34} < \lx < 10^{37}\ \ergs$.    Although the statistical measurement errors on each data point are typically $\approx$10\%, we adopt larger uncertainties here, to account for the radio and X-ray data not being strictly simultaneous, and to account for the lack of detailed spectral information for each data point.  Following  \citet{gallo14}, we adopt errors of 0.15 and 0.30 dex on each radio and X-ray luminosity, respectively.    \citet{corbel08} also re-examine strictly simultaneous \textit{Chandra} and VLA observations of \src\ in quiescence from 2003 July 28-29 (56 ks with \textit{Chandra} and 14 h with the VLA; also see \citealt{hynes04, hynes09}).  We adopt their X-ray and radio flux measurements, including a 3-9 keV X-ray flux of $1.79^{+0.13}_{-0.06} \times 10^{-13}\ \flux$ and a radio flux density of $0.193\pm0.022$ mJy at 8.4 GHz (they find a radio spectral index of $\alpha_{\rm r}=0.29\pm0.46$).  For placing these observations onto the radio/X-ray plane, the above fluxes correspond to X-ray and radio luminosities of $\lx = 2.8\times10^{32}\ \ergs$ (1-10 keV) and $\lr = 1.1 \times 10^{28}\ \ergs$ (8.4 GHz).  We adopt 30\% systematic uncertainties on these   X-ray and radio fluxes to account for variability in each band during the strictly simultaneous observations (0.13 dex).  

\subsubsection{\citet{rana16}}
\label{sec:obs:lit:r16}

\citet{rana16} present  VLA observations in quiescence taken on  2013 December 2, along with three epochs of \textit{NuSTAR} X-ray observations (2013 October 13, October 14, and December 2) and one epoch of \textit{XMM-Newton} observations (2013 October 13).  Here, we only consider their simultaneous radio and X-ray epochs from December 2, which includes 9 h with the VLA and 25 ks with \textit{NuSTAR}.

Their VLA observations are from 5--8 GHz,  and they provide radio flux density measurements at four central frequencies (each with 512 MHz bandwidth; see their Table~3).  To compare to  8.4 GHz flux densities from our 2015 campaign, we fit a power-law to their published 5--8 GHz radio spectrum, using the same routine described in Section~\ref{sec:obs:vlaspec}.  We find a radio spectral index $\alpha_{\rm r} = -0.27 \pm 0.05$ and $f_{\rm 8.4} = 0.274 \pm 0.082 $ mJy ($\lr = 1.6 \times 10^{28}\ \ergs $). 

For the X-ray flux coinciding with their VLA observation, we estimate an  absorbed 3-10 keV X-ray flux of $2.5 \times 10^{-13}~\flux$ during their December 2 \textit{NuSTAR} epoch (see their Figure 6).   Using their best-fit power-law spectrum to \src\ in quiescence ($\nh = 1.2 \times 10^{22}\ {\rm cm}^{-2}$ and $\Gamma=2.12 \pm 0.07$; 90\% confidence), we estimate $f_{\rm 1-10 keV} = 5.4 \times 10^{-13}~\flux$ ($\lx = 3.7 \times 10^{32}\ \ergs$).   The difference in $\nh$ obtained by \citet{rana16} compared to our best-fit $\nh$ in Section~\ref{sec:obs:xspec} is because \citeauthor{rana16}\ adopt \citet{wilms00} abundances during their fits, and we adopt \citet{anders89} abundances (see Appendix~\ref{sec:app:nh}).  We add 30\% systematic errors  (0.13 dex) to both the radio and X-ray luminosities to account for  variability.  

\subsubsection{Quiescent X-ray Spectra}
\label{sec:obs:lit:spec}

\citet{reynolds14} take a comprehensive look at four pre-outburst X-ray spectra, obtained by \textit{Chandra}, \textit{XMM-Newton}, and \textit{Suzaku}.  Over these four epochs, they find 0.3-10 keV X-ray fluxes ranging from $0.8 \times 10^{-12} - 3.4 \times 10^{-12}~\flux$, and $\Gamma$ ranging from $1.95 - 2.25$.  They also perform a joint spectral fit to these four spectra, forcing a common column density and powerlaw component, obtaining $\Gamma=2.05 \pm 0.07$ and $\nh = \left(1.15 \pm 0.07\right) \times 10^{22}~\cmtwo$ (90\% confidence).    We adopt $\Gamma=2.05$ as the `canonical' pre-outburst X-ray spectrum.   Although \citet{reynolds14} favor a larger column density than our study (likely because they adopt different abundances,  from \citealt{asplund09}),  adopting their $\Gamma=2.05$ will not systematically influence our conclusions.  Earlier studies based on the same pre-outburst \textit{Chandra} (two epochs) and \textit{XMM-Newton} observations (1 epoch)  quote best-fit column densities similar to our study --- $0.75_{-0.08}^{+0.15} \times 10^{22}$\ and $0.81\pm0.01 \times 10^{22}~\cmtwo$ from \textit{Chandra }\citep{corbel08}, and $0.88\pm0.6 \times 10^{22}~\cmtwo$\ from \textit{XMM-Newton} \citep{bradley07}.  Those earlier studies find photon indices of $\Gamma = 2.1 \pm 0.3$, $2.17\pm0.13$  and $2.09 \pm 0.08$, respectively, which are consistent with the range of $\Gamma$ in \citet{reynolds14}. 

We also include a \textit{Chandra} observation of \src\ obtained by \citet{tomsick15} on 2015 Nov 27 (obsID 17245; PI Tomsick; note that this was 26 days before \src\ flared again on 2015 Dec 23; \citealt{beardmore15a}).  This post-outburst observation shows X-ray properties similar to pre-outburst, including $\nh=\left(1.1 \pm 0.3\right) \times 10^{22}~\cmtwo$ (using \citealt{wilms00} abundances),  $\Gamma=2.0\pm0.3$ (90\% confidence), and an absorbed 0.3-10 keV flux of $7.6\times10^{-13}~\flux$.   When we display values from \citet{reynolds14} and \citet{tomsick15} in upcoming figures, we convert their reported 0.3-10 keV fluxes to 0.5-10 keV.

\section{Discussion}
\label{sec:disc}
 In Figure~\ref{fig:xrayevol} we display the X-ray spectral evolution of \src\ as a function of Eddington ratio.   This figure supports the statistical assertion by \citet{plotkin13} that the spectral softening occurs over a  narrow range of luminosity, before the \xrb\ reaches its minimum (i.e., pre-outburst) quiescent luminosity.  From Figure~\ref{fig:decay}, \src\ reaches $\Gamma \approx 2$ between \datefour\ - \dateseven, which indicates that \src\ re-enters  the quiescent spectral state around $L_{\rm 0.5-10 keV}  = 2.5 - 3.2 \times 10^{33}\ \ergs$ ($-5.6 \lesssim \log \lxledd \lesssim -5.5 $), and that the transition into quiescence occurs over only a factor of $\approx$3 in luminosity.  This luminosity where \src\ enters quiescence is lower than the 10$^{-5}~\ledd$ threshold suggested by \citet{plotkin13}, but it is still a factor of $\approx$3-4 above the average pre-outburst quiescent luminosity of $8 \times 10^{32}~\ergs$ for \src\ \citep[0.5-10 keV;][]{bernardini14}.  \citet{sivakoff15b} suggest that \src\ did not settle back to its pre-outburst $\lx$ until sometime between August 5 -- August 21 (interestingly,  the optical emitting outer disk did not return  to its pre-outburst flux level until 2015 October 10 - 20; \citealt{bernardini16b}).  
 
 The X-ray variability properties of \src\ during our campaign are also comparable to pre-outburst, as our maximum measured X-ray $\fvar = 55\pm 2\%$ is similar to $\fvar = 57.0 \pm 3.2$\% reported by \citet{bernardini14}.  We also demonstrate that the X-ray spectral softening is not accompanied by corresponding changes in the shape of the radio spectrum from the outer jet (\citealt{han92} also observed flat/inverted radio spectra at comparable radio flux densities during the decay of the 1989 outburst).  The normalization of the radio/X-ray luminosity correlation appears identical between the 2015 and 1989 outbursts, thereby suggesting a robust disk/jet coupling for \src.   The \xrb\ GX 339$-$4 also displays nearly identical  correlation slopes and normalizations between different outbursts \citep{corbel13}.

\begin{figure}
\begin{center}
\includegraphics[scale=0.37]{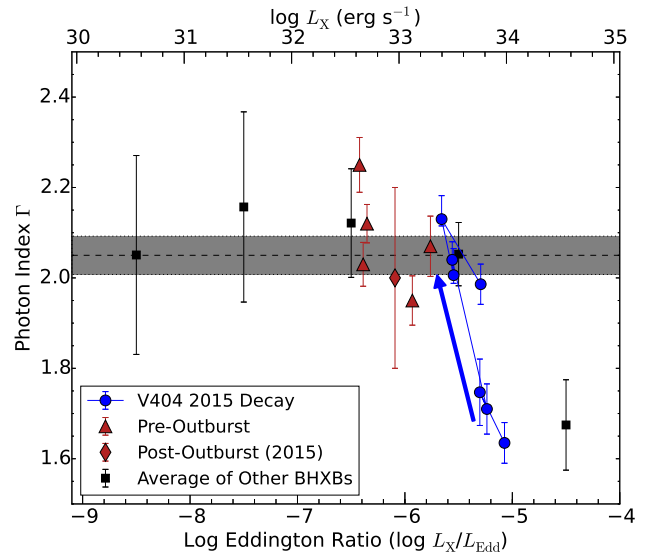}
\caption{The X-ray spectral evolution of \src\ as function of Eddington ratio ($\lxledd$, where $\lx$, which is displayed along the top axis,  is calculated from 0.5-10 keV).   Blue circles represent observations from our 2015 decay, with the blue solid lines connecting the data points to illustrate the evolution with time (following the direction of the blue arrow).  Red triangles show spectral fits from pre-outburst observations compiled from \citet{reynolds14} and \citet{rana16}, the red diamond represents a \textit{Chandra} observation from 2015 November \citep{tomsick15}, and black squares represent the average photon indices of ten quiescent \xrb s binned by Eddington ratio,  from \citet{plotkin13}.  The X-ray spectral softening occurs  over a narrow range in Eddington ratio (from $10^{-5.1}$ to 10$^{-5.6}$), which is more luminous than the average pre-outburst  Eddington ratio ($10^{-6.2}$; \citealt{bernardini14}). The dashed line and grey shaded region illustrate the average $\Gamma$ and 1$\sigma$ confidence interval 
for \src\ pre-outburst (from \citealt{reynolds14}).}
\label{fig:xrayevol}
\end{center}
\end{figure}

\subsection{Comments on RIAF X-ray Emission}
\label{sec:disc:riafxray}

Our data are consistent with a RIAF origin for the X-rays from \src, as long as the  X-ray emission is very inefficient throughout the entire decay as described below.\footnote{We note, however,  that there are arguments against a RIAF interpretation for \src\ in the literature, including a lack of  X-ray emission lines from \src\ in quiescence \citep[e.g.,][]{bradley07, rana16}, and also UV emission that is inconsistent with most  RIAF models unless an outflow is  incorporated  \citep{hynes09}.  } 
 We can parameterize the X-ray luminosity as $\lx \propto \Mdot^q$, where $q$ describes the radiative efficiency (for convenience, we will refer to $q$ as the `radiative efficiency') and $\Mdot$ is the mass accretion rate through the inner regions of the accretion flow.  For a partially self-absorbed synchrotron jet, the radio luminosity  follows $\lr \propto Q_{\rm j}^{17/12 - \left(2/3\right)\alpha_{\rm r}}$, where  the jet power, $Q_{\rm j}$, is linearly proportional to $\Mdot$ \citep[e.g.,][]{falcke95}.  We adopt $\alpha_{\rm r} = 0$, such that the slope of the radio/X-ray correlation $m=\left(17/12\right)/q$.  For \src, $m \approx 0.5$, which implies an X-ray efficiency of $q \approx 2.8$ (for the most inverted radio spectrum observed during our campaign, $\alpha_{\rm r} \approx 0.2$, the implied X-ray efficiency is $q\approx 2.6$, and an inverted radio spectrum does not alter our conclusions).  This efficiency is consistent with expectations from many RIAF models.  For example, \citet{merloni03} calculate that the X-ray efficiency may range from $q_{\rm RIAF} \approx 2.0-3.4$ at the lowest accretion rates.

 It is beyond the scope of this paper to explore specific RIAF models in detail.  However, any model must satisfy  other observational constraints besides $q \approx 2.8$.  One is the relatively rapid X-ray spectral softening.     In a hot accretion flow, synchrotron self-Compton (SSC) processes are important for generating X-ray emission  \citep[see, e.g., recent reviews by][]{poutanen14, malzac16}, and the X-ray spectral softening toward quiescence is generally expected  to be driven by a lower optical depth to inverse Compton scatterings and/or a lower flux of seed  photons as $\Mdot$ decreases \citep[e.g.,][]{esin97, tomsick01, sobolewska11, niedzwiecki14}.  However, for \src, our monitoring campaign demonstrates that such a decrease in optical depth/seed photon flux cannot be accompanied by a  large change in the the X-ray efficiency $q$, since the slope of the radio/X-ray correlation does not change at a detectable level (i.e., the uncertainty on the best-fit $m=0.54 \pm 0.03$ implies $\sigma_q \approx \pm 0.08$).   Furthermore, the X-ray variability properties of \src\ are similar across our entire campaign and pre-outburst, in both flux amplitude and timescale, which may indicate that the size of the X-ray emitting region does not evolve significantly.    
    
 \subsection{Comments on Jet Synchrotron X-ray Emission}   
 \label{sec:disc:jetxray}
 
 Several  studies of \src\ in quiescence have   favored a synchrotron origin for the X-ray emission \citep[e.g.,][]{bernardini14, xie14, markoff15}.    In this case, the observed $\Gamma\approx2$ implies that the synchrotron emitting particles are  radiatively cooled \citep[see, e.g.,][]{plotkin13}, and/or that the particle acceleration mechanisms along the jet become less efficient as luminosity decreases (i.e., the maximum Lorentz factor of accelerated particles becomes smaller, see, e.g.,\ Connors et al.\ subm. for recent discussions on both scenarios).   
 
  In the case of radiatively cooled particles,  the X-ray spectral softening implies  a switch in the X-rays from being dominated by the RIAF and/or by the optically-thin jet  in the hard state, to becoming dominated by a (synchrotron cooled) jet in quiescence \citep{yuan05}.    However, as described below, our 2015 campaign excludes  a synchrotron cooled jet in quiescence, unless the emission is scattered into the X-ray waveband through SSC.   For synchrotron cooled X-rays, the radiative efficiency $q_{\rm cool} = p + 2 - \left(3/2\right)\Gamma$ \citep{heinz04}, where $p$ describes the energy distribution of the synchrotron emitting particles \textit{before} they are cooled by radiative losses (i.e., below the cooling break, the number density of relativistic particles $n_e \propto \gamma ^{-p}$, where $\gamma$ is the Lorentz factor of the emitting particles).  For $2.0 < p < 2.3$ (which is typical in astrophysical contexts; e..g, \citealt{bell78, drury83, achterberg01}),  $\Gamma=2$ yields $1.0 < q_{\rm cool} <1.3$, which results in  a steeper  radio/X-ray correlation slope of $1.1 < m _{\rm cool}< 1.4$ (for an inverted  $\alpha_{\rm r}\approx 0.2$, the shallowest slope supported by our data would be $m_{\rm cool} \approx 1.0$).

If X-rays are to become synchrotron cooled in quiescence, then  the transition must occur at $\lx \approx 3 \times 10^{33}\ \ergs$ (i.e.,  where the X-ray spectrum reaches $\Gamma \approx 2$).   However, a steepening of the radio/X-ray correlation at that X-ray luminosity predicts a quiescent  radio luminosity that is $\approx$0.5-0.9 dex lower  than was observed in either 2003 or 2013 (for $p=2.0$-2.3; see Figure~\ref{fig:lrlxzoom}).  We note that empirical studies on the broadband SEDs of other hard state \xrb s suggest that the spectrum of optically thin jet  synchrotron emission could (on average) follow $f_\nu \propto \nu^{-(0.7 - 0.8)}$ \citep{russell13}, from which one  infers $p$ as large as 2.6, corresponding to $q_{\rm cool} =1.6$.  Synchrotron cooled X-ray emission could therefore yield a  slope  as shallow as $m_{\rm cool}=0.9$. Even in this limiting case, synchrotron cooled X-ray emission  underpredicts the observed pre-outburst radio luminosities of \src\ by $\approx$0.3 dex.    Synchrotron cooled X-rays in quiescence also appear unlikely from the radio properties of A0620$-$00 \citep{gallo06} and XTE~J1118+480  \citep[][both sources at $\lx \approx10^{-8.5}~\ledd$]{gallo14}.  Although, for those two sources it is not possible to isolate the  inflection point in X-ray luminosity where the radio/X-ray correlation should steepen.   
  
\begin{figure}
\begin{center}
\includegraphics[scale=0.45]{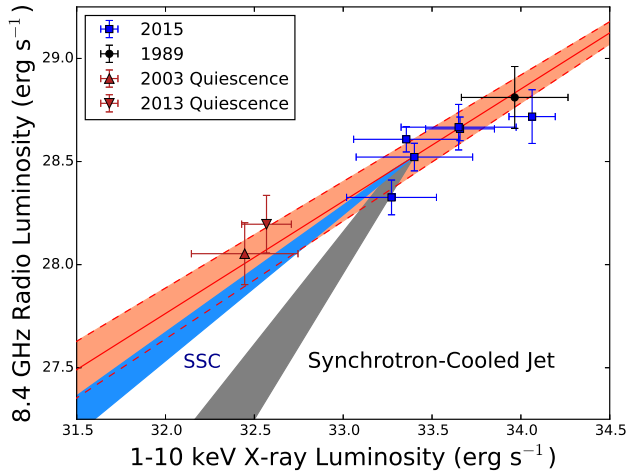}
\caption{The radio/X-ray luminosity correlation from  Figure~\ref{fig:lrlxfull}, zoomed in on the 2015 data.   Symbols have the same meaning as in Figure~\ref{fig:lrlxfull}.  The gray shaded region shows the predicted path \src\ would take through the radio/X-ray plane if the X-ray spectral softening signifies a switch to synchrotron-cooled X-ray emission in quiescence, a scenario that is inconsistent with the pre-outburst radio luminosity.  The light blue shaded region illustrates the path if the X-rays switch to SSC emission with synchrotron cooled seed photons.}
\label{fig:lrlxzoom}
\end{center}
\end{figure}

 In light of the above, we suggest two possibilities for a jet X-ray origin in quiescence: (1) the X-ray emission is SSC with synchrotron cooled seed photons (also see, e.g., \citealt{markoff05} for discussions on SSC from jet models); or (2) particle acceleration along the jet becomes less efficient with decreasing luminosity.  In the first case, for SSC arising from an optically thin plasma, $q$ is generally expected to be larger (i.e., less efficient) than the value for the mechanism that produces the source of seed photons,  on account of   SSC depending on the product of the photon field density and the particle density \citep[e.g.,][]{falcke95}.  For simplicity, we will assume here that $q$ will increase by 1 (e.g., for a conical jet without  velocity gradients in the bulk flow, the  particle density normalization is expected to scale linearly with $\Mdot$; \citealt{heinz03}).  In that case, the radiative efficiency of SSC (with synchrotron cooled seed photons) for $2.0 < p < 2.3$  would be  $2.0 < q_{\rm SSC, cool} < 2.3$, implying $0.6 < m_{\rm SSC, cool} < 0.7$ (or $m_{\rm SSC, cool} > 0.5$ for $p < 2.6$).  Such a change in slope of the radio/X-ray correlation (i.e., $m\approx 0.5$ in the hard state  to $0.6-0.7$ in quiescence, for $2.0 < p  < 2.3$) may not be detectable given the pre-outburst luminosity of \src\ (see Figure~\ref{fig:lrlxzoom}).   
 
   SSC X-rays with synchrotron cooled seed photons could also plausibly explain the softer quiescent X-ray spectrum, if the Comptonized spectrum is produced by single scatterings off an optically thin plasma.\footnote{As noted by \citet{corbel08}, the slope of $m \approx 0.5$ in the hard state could be consistent with SSC from \textit{ optically thin} synchrotron seed photons with $p=2.0-2.3$, which yields $q_{\rm SSC, thin} = 2.8$ and $m_{\rm SSC, thin} \approx 0.5$.  Although,  such an interpretation in the low-hard state is less likely for other \xrb s, since other sources generally show steeper radio/X-ray correlation slopes.}   To  properly assess the radiative efficiency and spectral shape of SSC emission would require detailed Componization modeling, to simulate the number of scatterings, the energy distribution of the synchrotron emitting particles, and adiabatic cooling losses related to the escape rate of the highest energy particles (also see, e.g., \citealt{malzac09} for  relevant discussions on calculating SSC in the context of a coronal \xrb\ model), which we intend to address in a future paper. 

   For the second possibility of inefficient particle acceleration, the observed $\Gamma \approx 2$ implies (uncooled) optically thin synchrotron radiation emitted by a non-thermal population of particles with $p \approx 3$, which results in  $q_{\rm thin} \approx \left(17/12\right) + \left(p-1\right)/3  = 2.1$  \citep[e.g.,][and references therein]{plotkin12}.  In this case we expect  a radio/X-ray correlation slope of $m_{\rm thin}=0.7$, consistent with our prediction for SSC emission.   The X-ray spectral softening during the transition into quiescence is straightforward to explain if the larger $\Gamma$ is driven by a transition to optically thin synchrotron radiation emitted by a population of non-thermal electrons with $p=3$, since $\Gamma = \left(p+1\right)/2$ for optically thin synchrotron.   We cannot distinguish between SSC and inefficient particle acceleration here.

Finally, we note that at even lower luminosities ($\lx \approx 10^{-8.5} \ledd$), a transition to SSC X-rays in quiescence has also been suggested from broadband modeling of  two other systems---A0620$-$00 \citep{gallo07} and XTE J1118+480 \citep{plotkin15}---and also  phenomenologically from the broadband spectrum of a third system, Swift J1357.2$-$0933 \citep{plotkin16}; this transition may also be accompanied by a decrease in the particle acceleration efficiency \citep[e.g.,][although see \citealt{markoff15}; Connors et al.\ subm.\ regarding degeneracies between weak particle acceleration and radiative cooling losses]{plotkin15}.  An important caveat, however, is that in the above  cases the source of synchrotron seed photons is proposed to be emitted by a mildly relativistic population of electrons following a thermal distribution of energies, and not necessarily synchrotron cooled emission from a non-thermal jet (see \citealt{shahbaz13, plotkin15}; Connors et al.\ subm.\ for details).

\subsection{A Tentative Size for the Compact Jet}
\label{sec:disc:jetsize}

In Section~\ref{sec:res:tdelay}, we present tentative evidence for correlated radio and X-ray emission  on \dateeight\  with a $\tdelay \pm \tdelayErr$ min radio lag.  The best-fit correlation slopes on \dateeight\ ($m=0.42 \pm 0.15$; or $m=0.59 \pm 0.17$ after removing the radio time delay) are furthermore consistent with a RIAF or jet origin (from SSC or weakly accelerated particles) for the X-ray emission.  If this correlated variability is real, then it  implies that disk/jet couplings hold on minute-long timescales, even in quiescence.  Previous long simultaneous radio and X-ray observations of \src\ in quiescence did not detect correlated radio and X-ray variations,  most likely because our study is the first with sufficiently matched radio and X-ray sensitivities:  the 2003 VLA and \textit{Chandra} campaign \citep{hynes04, hynes09} was performed before the VLA upgrade, so that radio light curves were binned on $\approx15-20$ min intervals; the more recent 2013 campaign (with the upgraded VLA and \textit{NuSTAR}; \citealt{rana16}) was limited to $\approx$50 min X-ray time bins due to the sensitivity of \textit{NuSTAR}.  Those two previous campaigns do, however, exclude the possibility of longer radio and X-ray time lags (up to $\approx$5-10 hours; the possibility of even longer lags is not constrained).  

The measurement of a radio time delay opens the possibility of placing constraints on the size of the radio jet.  If we denote $z$ as the distance from the black hole along the axis of the jet, then we can approximate the X-rays as originating at $z\approx 0$ (i.e., the X-rays are emitted very close to the black hole, at a location consistent with the base of the jet), and we can define the 8.4 GHz radio emission as originating from a region located at a larger distance $z_0$.  If information from X-ray variations propagate down the axis of the jet with a dimensionless bulk speed $\beta=v_{\rm b}/c$ (where $v_{\rm b}$ is the bulk speed and $c$ is the speed of light), then $z_0 = \beta c \Delta t \left(1 - \beta \cos \theta \right)^{-1}$, where $\Delta t$ is the radio time delay,  $\theta$ is the viewing angle between the jet axis and our line of sight, and $\left(1 - \beta \cos \theta\right)$ is a correction term related to superluminal motion.  By approximating $\theta$ as the orbital inclination  $i=67^{+3}_{-1}$ deg, then  $\beta<1$ places an  upper limit on the jet size to $z_0 < \jetsize \pm \jetsizeErr$ AU.   This size limit  is based only on geometric arguments and not specific to any jet model, the only assumptions being that the X-ray flare signifies the beginning of material propagating down the jet, the adopted viewing angle,  and that there is no velocity gradient in the bulk flow.  The latter two assumptions in particular require further scrutiny, however.  For example, VLBA observations during the 2015 outburst of \src\ suggest that assuming $\theta=67$ deg for the viewing angle might  not be valid  (Miller-Jones et al.\ in prep).

Despite  the above approximations, the calculated $z_0 < 3.0$ AU limit  is consistent with a direct limit placed on \src\ in quiescence, where the compact core remains unresolved in high-resolution radio observations, providing a projected angular size  $<$1.3 milliarcsec \citep{miller-jones08}, which corresponds to a physical length  $z_0<3.4$ AU (assuming $d=2.39$ kpc and $\theta=67^\circ$).   If the X-ray variations are indeed propagating down the jet toward the radio photosphere, then in addition to the time delay, we  expect the radio light curve to be a `smoothed' version of the X-ray light curve:  the radio emitting region will have a larger physical size than the X-ray emitting one, thereby smearing the radio signal (according to the light travel time across the radio emitting region) and supressing the highest (temporal) frequency variations \citep[see, e.g.,][and references therein]{gleissner04}.   If we smooth our X-ray light curve with a 10 min sliding filter\footnote{The appropriate  filter is jet model dependent, but it is expected to be of comparable (or smaller) length than the measured time delay.  For example, for a conical geometry, one  expects the size of the radio emitting region to be $z_0 \tan \phi$, where $\phi$ is the opening angle of the jet. }
, then the X-ray variations decrease from $\fvar=0.28\pm0.03$ (see Table~\ref{tab:fvar}) to $\fvar=0.23 \pm 0.03$, which is similar to the observed radio light curve from which we measure $\fvar=0.22 \pm 0.02$.  Although this is an intriguing result, further study is required.  For example, if we also reduce the amplitude of the X-ray variability by $\lr^{0.54}$ (i.e., according to the radio/X-ray luminosity correlation), then smoothing the X-ray light curve yields rms fluctuations  ($\fvar=0.15\pm0.03$) that are smaller than observed in the radio.  Plus, the above ignores the effects of photon pileup on the \textit{Chandra}-based $\fvar$ estimate (although these effects are expected to be small; see Appendix~\ref{sec:app:xray}).

We stress that our radio and X-ray observations on \dateeight\ do not cover an entire flare, thereby making it difficult to understand systematics on our radio lag measurement.  Thus, we present the $z_0 < \jetsize \pm \jetsizeErr$ AU limit here as an example of the type of constraints that are attainable with current facilities, if one were to obtain longer stretches of strictly simultaneous radio and X-ray coverage (and measuring a time delay at $\geq 2$ radio frequencies might provide knowledge of $\beta$, which would yield a measurement on $z_0$ instead of a limit.\footnote{We searched for a frequency dependent time delay by cross-correlating the \dateeight\ X-ray light curve to the radio light curves at each of our four observing frequencies (from 5.2-11.0 GHz).  Although delays were detected at all four frequencies, all  were consistent  with $-15 \pm 4$ min with no discernible trend with observing frequency, which implies that a larger range of radio frequencies should be searched, and/or that systematics related to the short time coverage are influencing our results.})  
  So far,  constraints on jet sizes are sparse, as, even at higher luminosities,  we have direct constraints on the sizes of the compact, partially self-absorbed radio core for only three sources that have been resolved in the radio: GRS 1915+105 \citep[projected size $\approx$25-30 AU at 8.4 GHz and 8.6 kpc source distance;][]{dhawan00, reid14}, Cyg X-1 \citep[projected size $\approx$28 AU at 8.4 GHz and 1.86 kpc source distance;][]{stirling01, reid11}, and MAXI J1836$-$194  \citep[projected size $\approx$60-150 AU at 2.3 GHz, albeit with  an uncertain distance of 4-10 kpc;][]{russell15}, plus the aforementioned limit on \src\ in quiescence.  A more precise measure for \src\ would open the door to comparative studies to study how physical properties evolve from the hard state to quiescence, providing crucial constraints to inform jet models \citep[see, e.g.,][]{heinz06}.

\section{Conclusions}
We have presented a series of X-ray and radio spectra of \src\ during the end of its 2015 outburst, as it  transitioned back into quiescence.   Even though \src\ was a factor of $\approx$3-12 ($\approx$2-5) more luminous in the X-ray (radio) during our campaign compared to  pre-outburst,  by our final observing epoch its other multiwavelength properties were similar to pre-outburst, including a soft X-ray spectrum ($\Gamma \approx 2$), modest X-ray variability ($\fvar \approx 20-50 \%$), and a flat/inverted radio spectrum.  We thus conclude that \src\ reached the quiescent spectral state before it settled to its minimum quiescent luminosity.

We suggest that \src\ enters the quiescent spectral state at $L_{\rm 0.5-10keV} \approx 3 \times 10^{33}~\ergs$  (determined by the luminosity when the X-ray spectrum finishes softening  to   $\Gamma \approx 2$).   There is no corresponding evolution in the shape of the radio spectrum, or in the slope of the radio/X-ray luminosity correlation.  From the latter, we exclude the possibility that X-ray emission is dominated by a synchrotron  jet in quiescence \citep{yuan05}, unless the X-rays are SSC with synchrotron cooled seed photons, or particle acceleration along the jet becomes less efficient in quiescence.  From  correlated X-ray and radio variability on our final observing epoch (\dateeight), we  tentatively measure the radio emission lagging behind the X-rays by $15.4\pm 4.0$ min, which would imply a jet size $<3.0 \pm 0.8$ AU (measured between the jet base and the location of the 8.4 GHz photosphere).  Better multiwavelength coverage of a simultaneous X-ray and radio flare in quiescence is required.

Because of its well-constrained distance and orbital parameters,   \src\ is an exceptionally important source for understanding quiescent accretion flows  and jets.    As a long-orbital period system with a large accretion disk, \src\ settles to a relatively high quiescent luminosity of $\approx$10$^{33}~\ergs$ \citep[e.g.,][]{menou99}.  In the future, it will be insightful to perform a similar campaign on a shorter orbital period system, to determine if \xrb s with smaller accretion disks (and lower minimum quiescent luminosities)  complete their X-ray spectral softening and re-enter the quiescent state as rapidly as \src, and/or at a similar luminosity.  Finally,  as the most luminous quiescent \xrb\ with a well-determined distance, our improved knowledge on \src\ in quiescence will help guide multiwavelength surveys  to discover new (and less biased) populations of \xrb s through their quiescent radiative signatures \citep[e.g.,][]{jonker11, strader12, chomiuk13, fender13, torres14,  miller-jones15, tetarenko16}.

\acknowledgments

We thank the anonymous referee for constructive comments that improved this manuscript.  We are especially grateful to the \textit{Chandra}, VLA, and \textit{Swift} teams for coordinating these observations, particularly Belinda Wilkes and Neil Gehrels for approving the \textit{Chandra} and \textit{Swift} observations, and especially  Scott Wolk for patiently assisting with multiple iterations of our \textit{Chandra} observing setup.  We also thank Mark Claussen, Gustaaf van Moorsel, and Heidi Medlin for their assistance with the VLA observations.    We thank Arash Bahramian for helpful discussions on the X-ray decay timescale that assisted the timing of our trigger, and Thomas Russell for assistance with the radio data reduction.  RMP acknowledges support from Curtin University through the Peter Curran Memorial Fellowship.  JCAM-J is supported by an Australian Research Council Future Fellowship (FT140101082).   GRS acknowledges support from NSERC Discovery Grants.  DA acknowledges support from the Royal Society.     Support for this work was provided by the National Aeronautics and Space Administration through Chandra Award Number  GO5-16032A  issued by the Chandra X-ray Observatory Center, which is operated by the Smithsonian Astrophysical Observatory for and on behalf of the National Aeronautics Space Administration under contract NAS8-03060.   The scientific results reported in this article are based to a significant degree on observations made by the \textit{Chandra} X-ray Observatory.  This research has made use of software provided by the Chandra X-ray Center (CXC) in the application packages CIAO.  The National Radio Astronomy Observatory is a facility of the National Science Foundation operated under cooperative agreement by Associated Universities, Inc.

\facilities{CXO, NuSTAR, Swift, VLA}

\software{CIAO (v4.8; \citealt{fruscione06}), HEASOFT, NUSTARDAS (v1.6.0), ISIS \citep{houck00}, CASA (v4.5; \citealt{mcmullin07})}


\appendix

\section{On the Column Density}
\label{sec:app:nh}

 Here, we address a couple points regarding our best-fit column density ($\nh = 8.4 \pm 0.2 \times 10^{21}~\cmtwo$).  First, as pointed out in the text, some studies of \src\ in quiescence favor a larger column density than us  (e.g., $1.2\times10^{22}~\cmtwo$ in \citealt{reynolds14} and \citealt{rana16}), which we attribute to using different abundances when modelling  photoelectric absorption.  To illustrate this point,   we refit our \dateeight\ \textit{Chandra} spectrum (i.e., our observation with the most counts) using \citet{wilms00} abundances, and we obtain $\nh= \left(1.2 \pm 0.1\right) \times 10^{22}\ \cmtwo$.  We stress that our conclusions on the X-ray spectral softening are not sensitive to which abundances are adopted (e.g., we find $\Gamma=1.94_{-0.02}^{+0.10}$ on \dateeight\ using \citealt{wilms00} abundances, compared to $\Gamma=1.99 \pm 0.04$ adopted in the text using  \citealt{anders89} abundances).
 
 Secondly,  throughout the text we adopt spectral parameters obtained from a  joint fit where we force a common $\nh$ across all epochs.   This decision is made because we do not see large fluctuations in $\nh$ during our campaign.   In light of the high column density observed during the first two weeks of the outburst \citep[e.g.,][]{motta16, sanchez-fernandez16}, we present more details here to justify our choice of using a common $\nh$.   In Figure~\ref{fig:confmap}, we display our best-fit $\nh$ and $\Gamma$ values when we jointly fit the \textit{Chandra} and \textit{Swift} spectra, but allowing both $\nh$ and $\Gamma$ to vary on each epoch.  There are some mild variations in $\nh$ between epochs, but these variations are consistent with the errors (and after taking into account the expected degeneracies between $\nh$ and $\Gamma$).  We   tend to obtain slightly higher best-fit $\nh$ values on epochs before \datefive\ ($> 10^{22}~\cmtwo$), which could be indicative of a lingering column that did not fully dissipate until  the final week of July.  However,  the differences pre- and post-\datefive\ are generally $<$1$\sigma$.   Furthermore,  \datefive\ is the date when we stopped using the \textit{Chandra} HETG  to mitigate pileup.  Thus, the slightly higher best-fit $\nh$ values earlier in the decay could feasibly be related to systematics caused by the poorer soft X-ray response, and we do not see overwhelming evidence for significantly varying column densities.   Finally, we display a spectral fit to the \dateeight\ \textit{Chandra} spectrum (using \citealt{anders89} abundances) in Figure~\ref{fig:chspec}  to illustrate the quality of our  fits.

\begin{figure*}
\begin{center}
\includegraphics[scale=0.6]{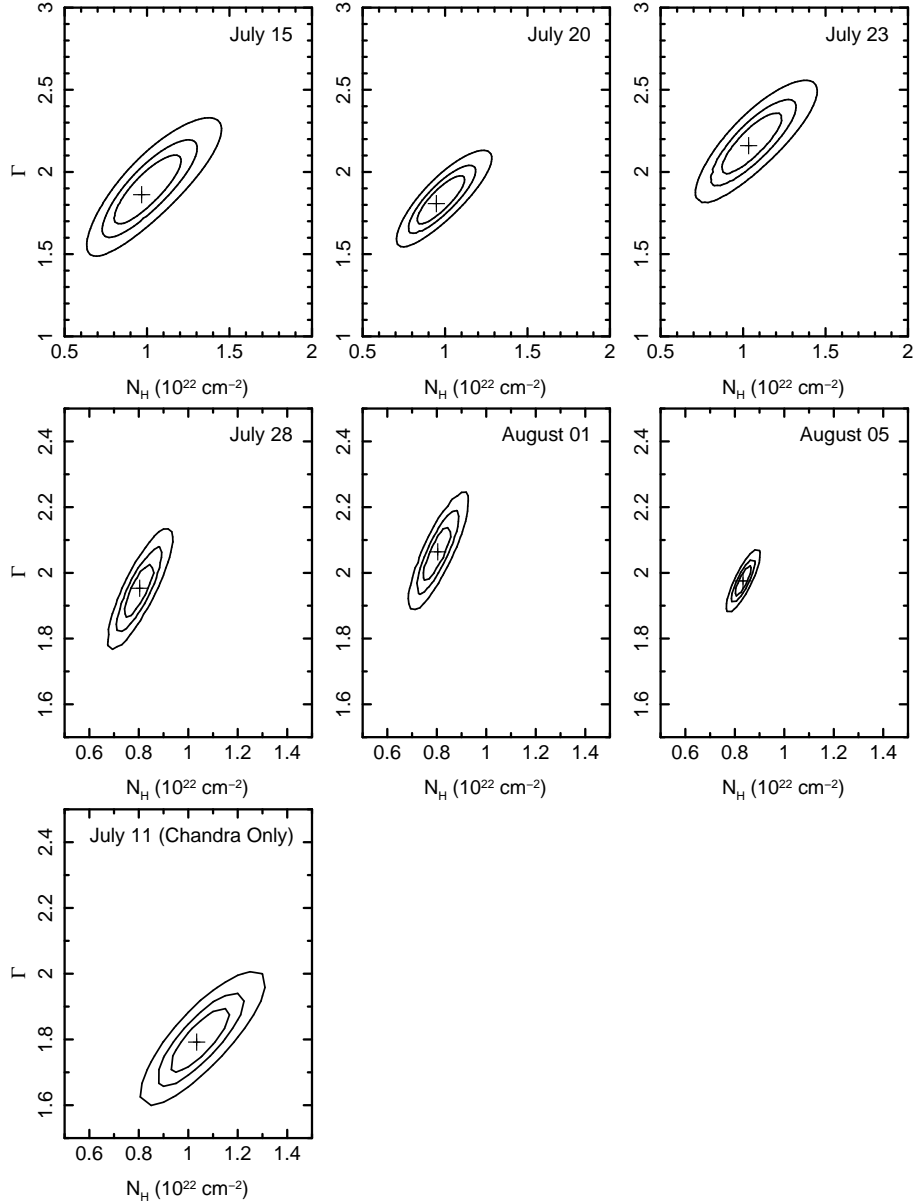}
\caption{Confidence contour maps of joint spectral fits to the \textit{Chandra} and \textit{Swift} observations for the photon index ($\Gamma$) and the column density ($\nh$), when allowing $\nh$ to vary as a free parameter on each epoch.  Contours denote 68, 90, and 99\% confidence intervals (corresponding to changes in the Cash statistic of $\Delta C = 2.3$, 4.6, and 9.2, respectively, for two parameters of interest).  The final panel for \dateone\ does not include \textit{Swift} data.   Note the different axis scales for the top row, and that because we allow $\nh$ to vary during each epoch, the errors illustrated here are larger than those listed in Table~\ref{tab:xrayspec}.}
\label{fig:confmap}
\end{center}
\end{figure*}

\begin{figure}
\begin{center}
\includegraphics[scale=0.6]{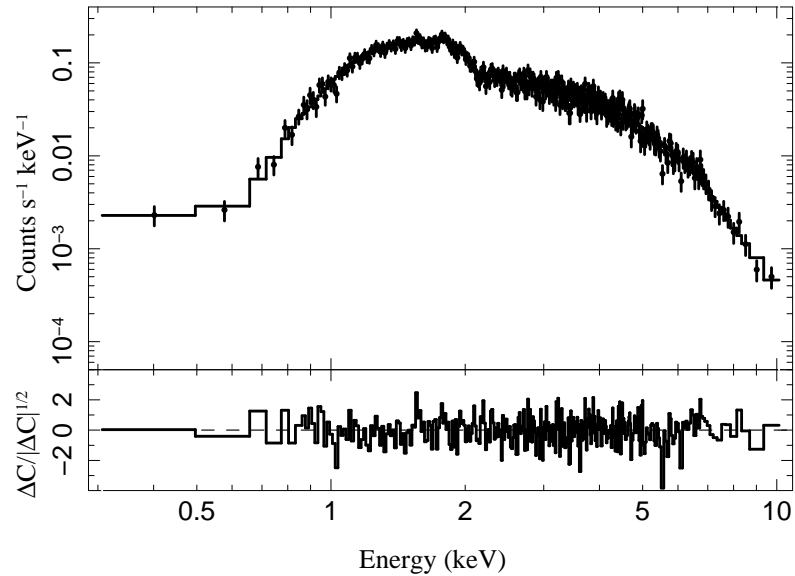}
\caption{\textit{Chandra} X-ray spectrum from \dateeight, as an example of the quality of our spectral fits.}
\label{fig:chspec}
\end{center}
\end{figure}

\section{Details on X-ray Spectral Analysis and Pileup Corrections}
\label{sec:app:xray}

Here, we expand on our discussion of the X-ray spectral analysis in Section~\ref{sec:obs:xspec} by describing tests regarding pileup corrections to the \textit{Chandra} observations.  One effect of pileup is \textit{energy migration}, where two photons are registered as a single event with an energy (improperly) set to the total energy of both photons, which causes the observed spectrum to appear harder than the intrinsic one.  By applying the \citet{davis01} pileup model to the \textit{Chandra} data during the spectral fits, we  correct for this effect and recover  unpiled photon indices.

One of the assumptions behind the \citet{davis01} model is that the  X-rays are emitted at a constant count rate.  However, as described in Section~\ref{sec:res},  \src\ was variable during the \textit{Chandra} observations.  To check that the variability was not strong enough to violate the assumption for constant count rate, we divided each \textit{Chandra} observation into 2-3 segments, with each segment filtered over time periods when \src\ displayed similar count rates (the boundaries for the count rate  filters were chosen so that each segment contained a comparable number of total counts).  For each epoch,  we then performed a joint fit to the 2-3 segments by allowing the photon index $\Gamma$ and the pileup grade migration parameter $\alpha$ to vary for each segment, but tying the best-fit column density $\nh$ to a common value.  We obtained similar best-fit photon indices for each segment, and the best-fit spectral parameters were comparable to the values obtained when we applied the pileup model to each full observation.   The only exception is on \datetwo, where the count rate from \src\ dropped to $<$0.03 count s$^{-1}$ for the final 400 s of the observation (compared to an average count rate of $0.065 \pm 0.004$ during the rest of the observation; at no other point in that observation was the count rate below $<$0.03 count s$^{-1}$).  While we do not expect that change in count rate to influence the pileup correction, it was not possible to empirically test this expectation since only eight counts were detected from \src\ over those 400 s.  We therefore  chose to remove those 400 s from the observation.  We cannot envision that this choice to err on the conservative side for the spectral fitting  would impact our other results, especially considering that we had \textit{Swift} monitoring on that date for over 19 hours, which provides additional variability information and indeed confirms that the flux level decreased shortly after the \datetwo\ \textit{Chandra} observation ended (see Figure~\ref{fig:xlc}). 

 Due to grade migration, photon pileup can also act to suppress the observed amplitude of  X-ray flux variations.  However, we do not find variability suppression to be a highly significant effect for our observations, based on the following test.   We filtered our \dateeight\ \textit{Chandra} spectrum (i.e., our brightest and most variable observation) to include only time periods with the highest count rates ($>$0.5 count s$^{-1}$).   The pileup fraction in this filtered spectrum remained mild at  14\% (with $\alpha \approx 0.96$).   Such a pileup fraction from observations with a 0.4 s frame time implies that the observed count rate is on average a factor 0.84 smaller than the average intrinsic count rate\footnote{see Equation 3 of \href{http://cxc.harvard.edu/ciao/download/doc/pileup\_abc.pdf}{http://cxc.harvard.edu/ciao/download/doc/pileup\_abc.pdf}.} 
 (the fractional variability will be reduced by a similar factor).  We stress that the above estimate is a limit, as we only  see such large count rates from $\approx$200-400 min into the \dateeight\ observation.  The suppression of variability from pileup during other less extreme X-ray flares will be less severe (including during the \dateeight\ flare from which we estimate a radio/X-ray time delay, which peaks at a \textit{Chandra} count rate $<$0.5 count s$^{-1}$).

As a final check on our pileup correction, we note that our final three observations show weak readout streaks, which are produced by photons that strike the detector while an ACIS frame is being read.  Only the readout streak from the final observation (obsID 16707; Aug 5) is strong enough for a meaningful spectral analysis (139 net counts, opposed to $\lesssim30$ net counts in each of the other two streaks).   For this final observation, we extracted a streak spectrum from 0.5-8.0 keV, using rectangular apertures aligned with the streak on either side of \src.  We created  rmf and arf files at the location of \src, using the tools {\tt mkacisrmf} and {\tt mkarf}, respectively.  The effective exposure time of the streak spectrum was 358 s, which is equal to $N_{\rm frames} * N_{\rm rows} * 4\times10^{-5}$ s, where $N_{\rm frames}$ is the number of frames read (total exposure time/frame time = 42696 s / 0.4 s), $N_{\rm row}$ is the number of rows in our streak apertures (84 rows), and $4\times 10^{-5}$ s is the readout time for each ACIS frame.  We fit an absorbed powerlaw model, finding $\nh=  1.0 \pm  0.4~\cmtwo$ and $\Gamma = 2.0 \pm 0.2$ when allowing $\nh$ to vary, and $\Gamma=1.8 \pm 0.2$ when freezing $\nh$ to $8.4 \times 10^{21} \cmtwo$.  Thus, the streak spectrum provides consistent results to the fits performed with the \citet{davis01} model (within the errors).


\end{document}